\documentclass[pra,aps,preprint,amsmath,amssymb]{revtex4}
\usepackage{graphicx}
\usepackage{epsfig}
\include{graphics}
\usepackage{epstopdf} 

\begin{document}

\title{Stabilizing optical solitons by frequency-dependent linear gain-loss 
and the collisional Raman frequency shift}

\author{Avner Peleg$^{1}$ and Debananda Chakraborty$^{2}$}
\affiliation{$^{1}$ Department of Mathematics, Azrieli College of Engineering, 
Jerusalem 9371207, Israel}
\affiliation{$^{2}$ Department of Mathematics, New Jersey City University, 
Jersey City, New Jersey 07305, USA}

\date{\today}

\begin{abstract} 
We study transmission stabilization of optical solitons against emission of radiation 
in nonlinear optical waveguides in the presence of weak linear gain-loss, cubic loss, 
and the collisional Raman frequency shift. We first show how the collisional Raman 
frequency shift perturbation arises in three different physical setups. We then show 
by numerical simulations with a perturbed nonlinear Schr\"odinger (NLS) model that 
transmission in waveguides with weak frequency-independent linear gain is unstable. 
The radiative instability is stronger than the radiative instabilities 
that were observed in earlier studies for soliton transmission in 
the presence of weak linear gain, cubic loss, and various 
frequency-shifting physical mechanisms. In particular, the Fourier spectrum of the 
radiation is significantly more spiky and broadband than the radiation's Fourier 
spectra in earlier studies. Moreover, we demonstrate by numerical simulations with 
another perturbed NLS model that transmission in waveguides with weak frequency-dependent 
linear gain-loss, cubic loss, and the collisional Raman frequency shift is stable. 
Despite the stronger radiative instability in the corresponding waveguide setup with 
weak linear gain, stabilization occurs via the same generic mechanism that was suggested 
in earlier studies. More precisely, the collisional Raman frequency shift experienced 
by the soliton leads to the separation of the soliton's and the radiation's Fourier 
spectra, while the frequency-dependent linear gain-loss leads to efficient suppression 
of radiation emission. Thus, our study demonstrates the robustness of the proposed generic 
soliton stabilization method, which is based on the interplay between perturbation-induced 
shifting of the soliton's frequency and frequency-dependent linear gain-loss. 
\end{abstract}

\maketitle

\section{Introduction}
\label{Introduction}

The cubic nonlinear Schr\"odinger (NLS) equation, which describes the dynamics of waves   
in the presence of second-order dispersion and cubic nonlinearity, is one of the most 
widely used nonlinear wave models in science and engineering. It was successfully 
applied to describe propagation of pulses of light in nonlinear optical waveguides \cite{Agrawal2019,Hasegawa95,Iannone98}, Bose-Einstein condensates \cite{Dalfovo99,BEC2008}, 
nonlinear waves in plasmas \cite{Malomed89,Asano69,Horton96}, and water wave dynamics 
\cite{Newell85,Ablowitz91,Osborne2010}. The fundamental NLS solitons are the most important 
solutions of the cubic NLS equation owing to their stability and shape preserving properties. 
Due to these properties, fundamental NLS solitons are being considered for applications 
in a wide array of nonlinear optical waveguide systems, including optical waveguide 
communication lines, pulse compression, pulsed waveguide lasers, and optical 
switches \cite{Agrawal2019,Hasegawa95,Iannone98,Mollenauer2006,Agrawal2020}. In what follows, 
we adopt the common terminology in nonlinear waveguide optics \cite{Agrawal2019,Hasegawa95}, 
and refer to the fundamental solitons of the cubic NLS equation as optical solitons.

In many of the applications, the usage of short optical pulses is highly advantageous. 
For example, the rates of transmission of information through optical waveguide 
communication lines can be significantly increased by employing short pulses 
of light \cite{Agrawal2019,Iannone98,Agrawal2020}. However, as the duration of the 
pulses is decreased, effects of additional physical processes, which can often be regarded 
as weak perturbations, become important \cite{Agrawal2019,Hasegawa95,Malomed89,Hasegawa87,
Agrawal2020}. Major examples for such perturbative processes are due to the effects of 
delayed Raman response \cite{Agrawal2019,Hasegawa87,Gordon86}, two-photon absorption 
\cite{Agrawal2007,Dekker2007,Borghi2017}, and third-order dispersion 
\cite{Agrawal2019,Kodama85,Elgin93,Elgin95}. When an optical soliton 
propagates in the presence of these weak perturbations, a variety of undesirable effects 
can arise. The most common undesirable effects induced by the weak perturbations 
are changes in the soliton's amplitude, frequency, position, and phase, and emission 
of radiation (i.e., emission of unlocalized small-amplitude waves) 
\cite{Agrawal2019,Hasegawa95,Malomed89,CCDG2003,Kaup90,PC2018}. The impact of these 
effects on the soliton is typically weak for short distances, but the cumulative impact 
over intermediate and long distances can be highly destructive. Furthermore, among these 
common effects, radiation emission is the most damaging. It can lead to pulse distortion 
and soliton destabilization even for single-soliton propagation \cite{Agrawal2019,PC2018,PC2023}. 
Additionally, in the case of single-sequence soliton transmission through an optical waveguide, 
the emitted radiation leads to long-range interaction between the solitons, which in turn causes 
the breakup of the soliton pattern \cite{CCDG2003,Smith89}. Moreover, in the case of multisequence 
soliton transmission through an optical waveguide, the radiation emitted by the solitons 
during frequent intersequence collisions can lead to severe pulse-pattern distortion 
and to destruction of the soliton sequences \cite{PNT2016,CPN2016,PNH2017,PNC2010}. 
For these reasons, many research efforts have been devoted to the investigation 
of the effects of radiation emission from NLS solitons \cite{Agrawal2019,Malomed89,Elgin95,CCDG2003,Kaup90,PC2018,PC2023,Smith89,Kaup95}, 
and to the development of methods for mitigating these effects 
\cite{Iannone98,Mollenauer2006,PC2018,PC2023,PNT2016,CPN2016,PNH2017,
Mollenauer97,MM98,Mecozzi91,Mollenauer92,Kodama92}.

One method for stabilizing soliton transmission against radiation emission and other 
unwanted effects, such as amplifier noise, is based on the application of guiding filters
\cite{Iannone98,Mollenauer2006,Mollenauer97,MM98}. This method was studied both theoretically 
and experimentally already in the 1990s \cite{Iannone98,Mollenauer2006,Mollenauer97,MM98,Mecozzi91,
Mollenauer92,Kodama92}. It was first realized that when the central frequency of 
the guiding filters is constant, the transmission is unstable against radiation 
emission \cite{Iannone98,MM98,Mollenauer92,PC2018}. The drastic improvement in transmission 
stability against radiation emission was achieved by the implementation of guiding filters 
with a central frequency that varies with propagation 
distance \cite{Mollenauer97,MM98,Mollenauer92,Kodama92}. 
This improvement has led to the experimental observation of stable soliton-based multisequence 
transmission over transoceanic distances with bit-error-rate values smaller 
than $10^{-9}$ \cite{MMN96}. In Refs. \cite{PC2018,PC2023}, we claimed and demonstrated 
that soliton stabilization by guiding filters with a varying central frequency is just one 
example for a more general mechanism for stabilizing NLS solitons against radiation 
emission effects. Furthermore, we proposed that the general mechanism of soliton stabilization 
is based on the interplay between perturbation-induced shifting of the soliton's frequency 
and frequency-dependent linear gain-loss. We demonstrated the proposed stabilization mechanism 
for the following two nonlinear waveguide systems. (1) A waveguide system, where the frequency 
shift is caused by the effects of delayed Raman response on single-soliton propagation (i.e., 
by the Raman self-frequency shift \cite{Gordon86}). (2) A waveguide system, where both the frequency 
shift and the frequency dependence of the gain-loss are due to guiding filters with a varying central 
frequency. We showed that in both systems, transmission stabilization is realized via the same 
physical mechanism. More precisely, the perturbation-induced shifting of the soliton's frequency 
causes the separation of the soliton's and the radiation's Fourier spectra, while the frequency-dependent 
linear gain-loss leads to efficient suppression of radiation emission effects \cite{PC2018,PC2023}.

Despite the progress achieved in previous studies on soliton stabilization against radiation 
emission effects, these earlier studies suffer from several important shortcomings. 
First, only two physical mechanisms for the soliton's frequency shift were considered, 
one that is due to the Raman self-frequency shift and another that is caused by guiding 
filters with a varying central frequency. Second, the underlying radiative instability 
that was suppressed by the stable optical waveguide setups in previous studies was moderate 
in the following sense. In the absence of one of the stabilizing factors (e.g., in waveguides 
with frequency-independent linear gain), the observed bandwidth of the destabilizing radiation 
was relatively narrow, and the observed pulse distortions did not contain any irregular features, 
such as highly spiky oscillations \cite{PC2018}. Thus, all the previous studies did not answer 
the important question regarding the applicability of the proposed general stabilization method 
in cases where the underlying radiative instability is strong (in the sense defined above). 
Third, the pulse distortions observed in the stable waveguide setups in previous studies 
\cite{PC2018,PC2023} were very different from the pulse distortions observed in Ref. 
\cite{PNT2016} at the onset of transmission instability of {\it highly stable} soliton-based 
multisequence transmission. Therefore, the physical mechanism leading to the pulse distortion 
and to the eventual instability in the latter transmission systems remains unknown.

In the current paper, we address the aforementioned important shortcomings of previous works 
on stabilization of optical solitons against radiation emission effects. For this purpose, 
we investigate soliton propagation in nonlinear optical waveguides in the presence of three 
perturbations due to weak linear gain-loss, cubic loss, and the collisional Raman frequency shift. 
We first explain how the collisional Raman frequency shift perturbation arises in three different 
nonlinear optical waveguide setups. We then investigate by numerical simulations with 
perturbed NLS models and by calculations with the adiabatic perturbation method the 
destabilization and stabilization of soliton transmission in the presence of the three 
weak perturbations.

In the case of waveguides with weak frequency-independent linear gain, our numerical simulations 
show that the transmission is unstable due to radiation emission. Furthermore, the radiative 
instability is stronger than the radiative instabilities that were observed in Ref. \cite{PC2018} 
for soliton propagation in the presence of weak linear gain, cubic loss, and various 
frequency-shifting physical processes. In particular, the pulse distortions and the radiation's 
spectrum are significantly more spiky than the pulse distortions and the radiation's spectra that 
were observed in Ref. \cite{PC2018}. Additionally, the radiation's spectrum has 
a significantly wider bandwidth compared with the radiation's spectra in Ref. \cite{PC2018}.  
Consequently, in the current study, the separation between the soliton's and the radiation's 
spectra is partial, whereas in Ref. \cite{PC2018}, the spectral separation was full.

In the case of waveguides with frequency-dependent linear gain-loss, our numerical simulations 
show that soliton transmission is stable. Furthermore, despite the stronger underlying radiative 
instability in the corresponding waveguide setup with weak frequency-independent linear gain, 
stabilization is achieved via the same general mechanism that was proposed in Refs. \cite{PC2018,PC2023}. 
More specifically, the collisional Raman frequency shift of the propagating soliton causes the 
separation of the soliton's and the radiation's Fourier spectra, while the frequency-dependent 
linear gain-loss leads to efficient suppression of radiation emission effects. Therefore, our study 
demonstrates the robustness and the general applicability of the proposed soliton stabilization method, 
which is based on the interplay between perturbation-induced shifting of the soliton's frequency 
and frequency-dependent linear gain-loss. We also point out that the weak pulse distortions observed 
in waveguides with weak frequency-dependent linear gain-loss, cubic loss, and the collisional Raman 
frequency shift are similar to the weak pulse-pattern distortions that were observed in Ref. \cite{PNT2016} 
at the onset of transmission instability of highly stable soliton-based multisequence transmission.     
This similarity indicates that radiation emission due to the collisional Raman frequency shift plays 
an important role in transmission destabilization in the highly stable soliton-based multisequence 
transmission setups, which were studied in Ref. \cite{PNT2016}.

Delayed Raman response is an important nonlinear perturbation in optical waveguides. 
It arises due to the finite time of nonlinear response of the waveguide's medium to the 
propagation of light \cite{Agrawal2019,Gordon86,Hasegawa87}. The main effect of delayed Raman 
response on single-soliton propagation in an optical waveguide is a continuous downshift 
of the soliton's frequency \cite{Agrawal2019,Gordon86,Hasegawa87,Mitschke86}. This effect,   
which is known as the Raman self-frequency shift, is a result of energy transfer from 
higher frequency components of the pulse to its lower frequency components \cite{Agrawal2019,
Gordon86,Hasegawa87}. The self-frequency shift is accompanied by emission of radiation 
\cite{Agrawal2019,Kaup95,PC2018}. As a nonlinear process, delayed Raman response also 
affects interpulse soliton collisions. Its main effects on a single fast two-soliton 
collision are an amplitude shift and a frequency shift \cite{Chi89,Malomed91,Kumar98,P2004,
CP2005,NP2010}, which are also accompanied by emission of radiation \cite{CP2005,NP2010}. 
We refer to these amplitude and frequency shifts as the collisional Raman amplitude shift 
and the collisional Raman frequency shift, respectively.

Raman-induced amplitude and frequency shifts in interpulse collisions can be beneficially 
used in a variety of applications, including amplification in optical waveguide communication 
lines \cite{Islam2004,Agrawal2005}, in tunable laser sources \cite{Agrawal2019,Stolen72}, 
and in supercontinuum generation \cite{Gordon89,Wabnitz2012}.   
However, these processes can also have very negative effects on pulse propagation in 
optical waveguide systems. An important example for the latter situation is provided by 
multisequence soliton-based on-off-keyed transmission systems, in which the information 
is encoded by soliton amplitudes. In particular, in Refs. \cite{P2004,CP2005,P2007,CP2008,PC2012}, 
we investigated the interplay between bit-pattern randomness and a variety of nonlinear and 
linear physical processes in these systems. We showed that the interplay between bit-pattern randomness 
and the collisional Raman amplitude and frequency shifts leads to high bit-error-rate values 
at intermediate and large transmission distances \cite{P2004,CP2005,P2007,CP2008,PC2012}. 
Moreover, the dominant mechanisms for transmission errors at these distances were associated 
with the collisional Raman amplitude and frequency shifts \cite{P2004,CP2005,P2007,CP2008,PC2012}.

We choose to study soliton transmission in the presence of linear gain-loss and cubic loss, 
since these perturbations are very common in many optical waveguide systems, and are typically 
the dominant dissipative processes in these systems. In this respect, linear gain-loss and 
cubic loss are central examples for dissipative perturbations in optical systems. 
The waveguide's cubic loss arises due to two-photon absorption (2PA) or gain-loss saturation \cite{Agrawal2007,Dekker2007,Borghi2017,Tanabe2022}. Propagation of optical pulses in the 
presence of cubic loss has been studied extensively, both in weakly perturbed linear 
waveguides \cite{Cohen2004,Cohen2005B,PNH2017B,Betz2021,NHP2022}, and in nonlinear 
waveguides \cite{Malomed89,Stegeman89,Silberberg90,Tsoy2001,PNC2010,Gaeta2012,
Malomed2014,PC2018,PC2020,NH2023,PC2023}. The subject received even more attention in recent 
years due to the importance of 2PA in silicon nanowaveguides, which play a key role in many 
applications in optoelectronic devices \cite{Agrawal2007,Dekker2007,Borghi2017,Gaeta2008}.

The other sections in the paper are organized in the following manner. In Section \ref{model},  
we present the basic perturbed NLS propagation model, and show how the collisional Raman frequency 
shift perturbation arises in three different nonlinear optical waveguide setups. 
In Section \ref{simu}, we study soliton dynamics in nonlinear waveguides in the presence of weak 
linear gain-loss, cubic loss, and the collisional Raman frequency shift. We present the results 
for pulse dynamics in the presence of frequency-independent linear gain in Section \ref{simu2}, 
and the results for pulse dynamics in the presence of frequency-dependent linear gain-loss in 
Section \ref{simu3}. Our conclusions are presented in Section \ref{conclusions}.  
In Appendix \ref{appendA}, we describe the methods that were used to determine transmission 
stability from the numerical simulations results.

\section{The basic nonlinear propagation model}
\label{model} 

\subsection{Introduction}
\label{model1} 
We consider propagation of short pulses of light in a nonlinear optical 
waveguide with second-order dispersion, cubic nonlinearity, and three 
weak perturbations due to frequency-independent linear gain, cubic loss, 
and collisional Raman frequency shift effects. The linear gain can be realized by 
distributed Raman amplification \cite{Islam2004,Agrawal2005}, and the 
cubic loss arises due to 2PA or gain-loss saturation \cite{Agrawal2007,Dekker2007,
Borghi2017,Tanabe2022}. The collisional Raman frequency shift effects, 
which were studied in Refs. \cite{CP2005,P2007,CP2008,PC2012}, will be briefly 
described in Subsection \ref{model2}. Additional physical setups that give rise 
to similar effects will be described in Subsections \ref{model3} and \ref{model4}. 
We denote the dimensionless envelope of the electric field by $\psi$, and the 
dimensionless distance and time by $z$ and $t$. The propagation is then 
described by the following weakly perturbed NLS equation \cite{basic_model}:               
\begin{eqnarray}&&
i\partial_z\psi+\partial_t^2\psi+2|\psi|^2\psi=
ig_{0}\psi/2 - i\epsilon_{3}|\psi|^2\psi
+\epsilon_{s}\exp\left[i\chi(t,z)\right]\partial_{t}|\psi|,
\label{cre1}
\end{eqnarray}  
where the dimensionless linear gain, cubic loss, and collisional Raman frequency 
shift coefficients, $g_{0}$, $\epsilon_{3}$, and $\epsilon_{s}$, satisfy $0 < g_{0} \ll 1$, 
$0 < \epsilon_{3} \ll 1$, and $0 < |\epsilon_{s}| \ll 1$. The real-valued function 
$\chi(t,z)$ is the phase-factor of the optical pulse. The second and third terms on 
the left hand side of the equation describe the effects of second-order dispersion 
and cubic nonlinearity. The first and second terms on the right hand side are due to 
weak frequency-independent linear gain and cubic loss. The third term is associated with 
weak collisional Raman frequency shift effects, and can also arise due to other effects, 
as described in Subsections \ref{model3} and \ref{model4}.

The dimensionless physical quantities in Eq. (\ref{cre1}) are related to the dimensional 
quantities by the standard scaling laws for NLS solitons \cite{Agrawal2019}. The same 
scaling laws were used in our previous works in Refs. \cite{PC2018,PC2023,PC2020}. 
More specifically, the dimensionless distance is 
$z=(|\tilde\beta_{2}|X)/(2\tau_{0}^{2})$, where $X$ is the dimensional distance, 
$\tau_{0}$ is the soliton's temporal width, and $\tilde\beta_{2}$ is the second-order
dispersion coefficient. The dimensionless time is $t=\tau/\tau_{0}$, where $\tau$ is the 
dimensional time. $\psi=(\gamma_{3} \tau_{0}^{2}/|\tilde\beta_{2}|)^{1/2}E$, where $E$ 
is the electric field and $\gamma_{3}$ is the cubic nonlinearity coefficient. The 
coefficients $g_{0}$ and $\epsilon_{3}$ are related to the dimensional linear gain and 
cubic loss coefficients $\rho_{1}$ and $\rho_{3}$ by $g_{0}=2\rho_{1}\tau_{0}^{2}/|\tilde\beta_{2}|$  
and $\epsilon_{3}=2\rho_{3}/\gamma_{3}$. The coefficient $\epsilon_{s}$ is related to 
the dimensional collisional Raman frequency shift coefficient $\tilde\epsilon_{s}$ by 
$\epsilon_{s}=2\tilde\epsilon_{s}\tau_{0}/|\tilde\beta_{2}|$. 
Specific expressions for $\epsilon_{s}$ in different physical systems will be given in 
Subsections \ref{model2}-\ref{model4}.

In the current paper, we study stabilization and destabilization of fundamental NLS solitons 
(optical solitons) in the presence of the collisional Raman frequency shift. The fundamental 
soliton solution of the unperturbed NLS equation is given by 
\begin{eqnarray} 
\psi_{s}(t,z)\!=\! \Psi_{s}(x)\exp(i\chi) = \eta\exp(i\chi)/\cosh(x),
\label{cre2}
\end{eqnarray}
where $x=\eta\left(t-y+2\beta z\right)$, $\chi=\alpha-\beta(t-y)+\left(\eta^2-\beta^{2}\right)z$, 
and $\eta$, $\beta$, $y$, and $\alpha$ are the soliton's amplitude, frequency, position, and phase.   
The parameter $\beta$ is also related to the soliton's group velocity.

\subsection{Derivation of the $\epsilon_{s}e^{i\chi}\partial_{t}|\psi|$ term  
in multisequence soliton-based transmission}
\label{model2} 

In this subsection, we derive the perturbation term 
$\epsilon_{s}\exp\left[i\chi(t,z)\right]\partial_{t}|\psi|$ for multisequence 
soliton-based transmission in optical fibers. The presentation briefly follows 
the main steps in the detailed derivation that was presented in Refs. 
\cite{CP2008,PC2012}.

We consider a multisequence soliton-based transmission system with $2N+1$ pulse sequences, 
and with constant frequency spacing $\Delta\beta$ between adjacent sequences. We assume 
that the number of sequences is large, such that $N \gg 1$, and that $\Delta\beta \gg 1$, 
which is the typical situation in massive multisequence soliton-based transmission 
\cite{MM98,Nakazawa2000,PNH2017B}. Following Refs. \cite{CP2008,PC2012}, 
we also assume on-off-keyed transmission, in which the information 
is encoded in pulse amplitudes. In this case, each pulse is located at the 
center of a time slot of width $T$, which is allocated for a single bit. The occupation 
of a time slot can be described by a binary random variable, which attains the value 1 
with probability $s$ if the time slot is occupied, and the value of 0 with probability 
$1-s$ if the time slot is empty. Thus, the mean fraction of occupied time slots in the 
system is $s$.

To derive the perturbation term, we first analyze the effects of a single fast 
two-soliton collision in the presence of the relevant physical perturbations, 
which are typically weak \cite{CP2008,PC2012}. Since we consider massive 
multisequence transmission, where $N \gg 1$, delayed Raman response is 
the dominant physical perturbation. 
Therefore, the perturbation that we consider is due to delayed Raman response.  
Using the results of the single-collision analysis, we derive a perturbed NLS 
equation that describes the propagation of a soliton in a given sequence under 
many collisions with solitons from all other sequences. In the current paper, 
we study soliton stabilization in the presence of the collisional Raman frequency 
shift, which is the dominant frequency shift in massive soliton-based multisequence 
transmission. Therefore, unlike the derivations in Refs. \cite{CP2008,PC2012}, 
which considered all the effects associated with delayed Raman response in the system, 
in the current derivation, we focus attention on the term that is associated with 
the dominant frequency shift, namely, the collisional Raman frequency shift. 
We also point out that the perturbed NLS equation is derived within a mean-field 
approximation, in which we assume that a soliton in a given sequence ``sees'' the 
amplitudes of the solitons in the other sequences as constant \cite{CP2005,CP2008,P2007,PC2012}.

We first consider a fast collision between a soliton in the $j$th sequence 
and a soliton in the $j'$th sequence in the presence of weak delayed Raman 
response. The collision-induced change in the electric filed of the $j$th 
sequence soliton, which is associated with the momentum exchange in the 
collision and with the collisional Raman frequency shift, is given 
by \cite{Kumar98,P2004,CP2005,CP2008,PC2012}: 
\begin{eqnarray} &&
\Delta\psi_{j}^{(c)}(t,z)\!=
-\frac{4i\epsilon_{R}\eta_{j'}}{|\beta_{j'}-\beta_{j}|}
\exp\left[i\chi_{j}(t,z)\right]\partial_t |\psi_{j}(t,z)|.
\label{cre3}
\end{eqnarray}        
In Eq. (\ref{cre3}), $0 < \epsilon_{R} \ll 1$ is the Raman coefficient, 
$\beta_{j}$, $\psi_{j}$, and $\chi_{j}$ are the frequency, electric field, 
and phase-factor of the $j$th sequence soliton at the collision distance, 
and $\eta_{j'}$ and $\beta_{j'}$ are the amplitude and frequency of the 
$j'$th sequence soliton at the collision distance \cite{CP2008,PC2012}.

We now use Eq. (\ref{cre3}) to derive the $\epsilon_{s}\exp\left[i\chi(t,z)\right]
\partial_{t}|\psi|$ perturbation term in the NLS equation, which describes the 
propagation of a soliton in the $j$th sequence under collisions with solitons 
from all other $2N$ sequences. For this purpose, we denote by 
$\Delta z_{c}^{(1)}=T/(2\Delta\beta)$ the distance traveled by the $j$th sequence 
soliton while passing two successive time slots in the $j-1$ or $j+1$ sequences. 
We then sum Eq. (\ref{cre3}) over all collisions occurring within the interval 
$\Delta z_{c}^{(1)}$ \cite{CP2008,PC2012}. We also use the relation 
$|\beta_{j'}-\beta_{j}|=|j'-j|\Delta\beta$, and the mean-field assumption, 
which implies $\eta_{j'}=\mbox{const}=1$ \cite{CP2008,PC2012}. This calculation yields 
the following expression for the total collision-induced change in $\psi_{j}$ 
in the interval $\Delta z_{c}^{(1)}$, $\Delta\psi_{j}^{(c)(tot)}$ \cite{CP2008,PC2012}: 
\begin{eqnarray} &&
\Delta\psi_{j}^{(c)(tot)}(t,z)\!=
-\frac{8iNs\epsilon_{R}}{\Delta\beta}
\exp\left[i\chi_{j}(t,z)\right]\partial_t |\psi_{j}(t,z)|.
\label{cre4}
\end{eqnarray}                           
The rate of change of $\psi_{j}$ due to the collisions is 
$\Delta\psi_{j}^{(c)(tot)}/\Delta z_{c}^{(1)}$. As a result, the perturbation term 
$S_{j}$ is \cite{CP2008,PC2012}:  
\begin{eqnarray} &&
S_{j}\!= i\frac{\Delta\psi_{j}^{(c)(tot)}(t,z)}{\Delta z_{c}^{(1)}} = 
\frac{16Ns\epsilon_{R}}{T}
\exp\left[i\chi_{j}(t,z)\right]\partial_t |\psi_{j}(t,z)|.
\label{cre5}
\end{eqnarray}      
We see that the perturbation term has the same form for all the soliton sequences. 
It follows that in massive soliton-based multisequence transmission, the perturbation 
term that is associated with the collisional Raman frequency shift is indeed of the 
form $\epsilon_{s}\exp\left[i\chi(t,z)\right]\partial_{t}|\psi|$. Furthermore, the 
coefficient $\epsilon_{s}$ is given by $\epsilon_{s}=16Ns\epsilon_{R}/T$, and is 
independent of the sequence index $j$. Additionally, since $N \gg 1$, the effects of 
the collisional Raman frequency shift term are significantly stronger than the effects 
of the Raman self-frequency shift term, $\epsilon_{R}\psi\partial_{t}|\psi|^2$ 
\cite{CP2008,PC2012}.

\subsection{Derivation of the $\epsilon_{s}e^{i\chi}\partial_{t}|\psi|$ term  
in waveguides with localized variations in the linear gain-loss coefficient}
\label{model3}  

As shown in Subsection \ref{model2}, the $\epsilon_{s}\exp\left[i\chi(t,z)\right]
\partial_{t}|\psi|$ perturbation term plays an important role in the description 
of soliton propagation in massive multisequence optical waveguide transmission systems. 
However, the physical relevance of this perturbation is certainly not limited to 
the latter systems. In the current subsection, we show that the same perturbation 
term appears in the description of soliton propagation in nonlinear optical waveguides 
with weak localized variations in the linear gain or linear loss coefficient.

We consider propagation of a single soliton in a nonlinear optical waveguide 
with weak localized changes in the linear gain or linear loss coefficient. 
We assume that the soliton propagates with velocity $2\beta_{0}$, 
where $\beta_{0}$ is a constant, relative to the gain-loss variations, 
and emphasize that the treatment can be extended to the case 
where $\beta_{0}$ is a slowly varying function of $z$. Similar to the 
analysis in Subsection \ref{model2}, we perform the derivation in two steps. 
In the first step, we analyze the effects of a single encounter of the soliton 
with a weak localized gain-loss variation. In the second step, we use the results 
of the single-encounter analysis to derive a perturbed NLS equation, which describes  
soliton propagation in the presence of a periodic sequence of weak localized variations 
in the linear gain or linear loss coefficient. We then show that the term 
$\epsilon_{s}\exp\left[i\chi(t,z)\right]\partial_{t}|\psi|$ appears in the latter 
perturbed NLS equation.

Let us analyze the effects of one encounter of the soliton with a weak variation 
in the linear gain or linear loss coefficient, which is located at $y_{b}$. 
The single-encounter dynamics can be described by the following 
weakly perturbed NLS equation: 
\begin{eqnarray}&&
i\partial_z\psi+\partial_t^2\psi+2|\psi|^2\psi=
i\epsilon_{1}\bar{c}h\left(\bar{a}(t-y_{b})\right)\psi/2,
\label{cre6}
\end{eqnarray}  
where $0 < |\epsilon_{1}| \ll 1$, $\bar{a}$, $\bar{c}$, and $y_{b}$ are constants,  
and where the function $h(\tilde{x})$ is symmetric and real-valued. We assume 
that $|\beta_{0}| \gg 1$, such that the soliton encounter with the gain-loss 
variation is fast. Additionally, we assume that $\bar{c}$ is of order $\beta_{0}$ 
and that $0 < |\epsilon_{1}||\bar{c}| \ll 1$. The initial condition for the 
single-encounter problem is a fundamental NLS soliton of the form (\ref{cre2}) with amplitude 
$\eta(0)$, frequency $\beta(0)=-\beta_{0}$, position $y(0)=y_{0}$, and phase $\alpha(0)=0$.         
Therefore, the soliton's location is $y_{s}(z)=y_{0}+2\beta_{0}z$.

It is convenient to carry out the perturbation analysis for the single-encounter 
problem in a reference frame, which is moving with velocity $2\beta_{0}$. 
In this new reference frame, the soliton is located at $y'_{s}(z)=y_{0}$, 
and the linear gain-loss variation is at $y'_{p}(z)=y_{b}-2\beta_{0}z$. 
Consequently, in the new reference frame, the single-encounter dynamics 
is described by the equation  
\begin{eqnarray}&&
i\partial_z\psi+\partial_t^2\psi+2|\psi|^2\psi=
i\epsilon_{1}\bar{c}h\left(\bar{a}(t-y_{b}+2\beta_{0}z)\right)\psi/2,  
\label{cre7}
\end{eqnarray}        
with the initial condition (\ref{cre2}) and initial parameter values $\eta(0)$, 
$\beta(0)=0$, $y(0)=y_{0}$, and $\alpha(0)=0$ \cite{similar_calculations}. 
Under the assumptions $|\beta_{0}| \gg 1$ and $0 < |\epsilon_{1}||\bar{c}| \ll 1$,  
we can employ a singular perturbation method similar to the one applied for 
analyzing fast two-soliton collisions in the presence of weak perturbations 
\cite{PCG2003,SP2004,CP2005,PNC2010,PC2020}, and also fast collisions between pulses 
of weakly perturbed linear evolution models \cite{PNH2017B,NHP2022}. Since the formal 
calculations are similar to the calculations carried out for fast two-soliton collisions 
in Refs. \cite{PNC2010} and \cite{CP2005}, we give here only the main steps and main 
results of the new calculations.

Within the framework of the singular perturbation method, we look for a solution of 
the perturbed NLS equation (\ref{cre7}) in the form 
\begin{eqnarray}&&
\psi(t,z)=\psi_{s}(t,z)+\phi(t,z), 
\label{cre8}
\end{eqnarray}           
where $\psi_{s}$ is the soliton solution (\ref{cre2}) of the unperturbed NLS equation, 
and $\phi$ describes the effects of the interaction with the weak localized gain-loss 
perturbation. We define $\Phi(t,z)=\phi(t,z)\exp[-i\chi(t,z)]$, where $\chi(t,z)$ is 
the soliton's phase-factor, and expand $\Phi(t,z)$ in a perturbation series with respect 
to the two small parameters $\epsilon_{1}$ and $1/|\beta_{0}|$ \cite{similar_calculations}.       
The perturbation series takes the form: 
\begin{eqnarray} &&
\Phi(t,z)=\Phi_{1}^{(0)}(t,z)+\Phi_{1}^{(1)}(t,z)+\Phi_{2}^{(0)}(t,z)
+\Phi_{2}^{(1)}(t,z)+\Phi_{2}^{(2)}(t,z)+\dots,
\label{cre9}
\end{eqnarray}     
where the subscript in each term on the right hand side stands for the combined order 
with respect to both $\epsilon_{1}$ and $1/|\beta_{0}|$, and the superscript is the 
order with respect to $\epsilon_{1}$.  
Substitution of the expansion (\ref{cre9}) into Eq. (\ref{cre7}) shows that the 
leading order of the effects of the encounter between the soliton and the gain-loss 
perturbation is order $\epsilon_{1}$. The equation for the dynamics of $\Phi$ 
in this order is \cite{similar_calculations}: 
\begin{eqnarray} &&
\partial_{z}\Phi_{1}^{(1)}(t,z)=
\epsilon_{1}\bar{c}h\left(x_{b}\right)\Psi_{s}(x), 
\label{cre10}
\end{eqnarray}      
where $x_{b}=\bar{a}(t-y_{b}+2\beta_{0}z)$. Integration with respect to $z$ yields 
\begin{eqnarray} &&
\Phi_{1}^{(1)}(t,z)=\frac{\epsilon_{1}\bar{c}\Psi_{s}(x)}{4\bar{a}\beta_{0}}
H\left(x_{b}\right), 
\label{cre11}
\end{eqnarray}     
where $H(x_{b}) \equiv \int dx'_{b}h(x'_{b})$ is an odd function of $x_{b}$. 
The total change in $\psi$ in order $\epsilon_{1}$ is obtained by integrating 
Eq. (\ref{cre10}) from $-\infty$ to $\infty$. This calculation 
yields \cite{similar_calculations}:      
\begin{eqnarray} &&
\Delta\Phi_{1}^{(1)}(x)=
\frac{\epsilon_{1}\bar{c} \tilde{h}}{4\bar{a}|\beta_{0}|}\Psi_{s}(x), 
\label{cre12}
\end{eqnarray}     
where $\tilde{h} = \int_{-\infty}^{\infty} dx_{b}h(x_{b})$. It is straightforward 
to show that the collision-induced change $\Delta\Phi_{1}^{(1)}(x)$ is associated 
with a collision-induced amplitude shift $\Delta\eta^{(c)}=\epsilon_{1}\bar{c} 
\tilde{h}\eta(0)/(2\bar{a}|\beta_{0}|)$, which is accompanied by emission of 
radiation \cite{similar_calculations}.

The other important effects of the soliton-perturbation encounter are of order 
$\epsilon_{1}/|\beta_{0}|$. The equation for the dynamics of $\Phi$ in this order 
is \cite{similar_calculations}:  
\begin{eqnarray} &&
i\partial_z\Phi_{2}^{(1)}=
-\left[(\partial_{t}^{2}-\eta^{2}(0))\Phi_{1}^{(1)}
+4\Psi_{s}^2\Phi_{1}^{(1)}+2\Psi_{s}^{2}\Phi_{1}^{(1)\ast}\right]
+i\epsilon_{1}\bar{c}h\left(x_{b}\right)\Phi_{1}^{(1)}/2.
\label{cre13}
\end{eqnarray}             
Since $\Phi_{1}^{(1)}(t,z)$ is an odd function of $z$, the only term on the right 
hand side of Eq. (\ref{cre13}), which does not vanish upon integration with respect 
to $z$ is the dispersion term $-\partial_t^2\Phi_{1}^{(1)}$. Therefore, substitution 
of relation (\ref{cre11}) into Eq. (\ref{cre13}) and integration over $z$ from 
$-\infty$ to $\infty$ yield \cite{similar_calculations}:    
\begin{eqnarray} &&
\Delta\Phi_{2}^{(1)}(x)=
\frac{i\epsilon_{1}\bar{c} \tilde{h}}{4\bar{a}|\beta_{0}|\beta_{0}}
\partial_{t} |\psi_{s}(x)|. 
\label{cre14}
\end{eqnarray}   
The collisional change $\Delta\Phi_{2}^{(1)}(x)$ is associated with a collisional 
frequency shift  
\begin{eqnarray} &&
\Delta\beta^{(c)}=
-\frac{\epsilon_{1}\bar{c}\tilde{h}\eta^{2}(0)}{6\bar{a}|\beta_{0}|\beta_{0}},  
\label{cre15}
\end{eqnarray}   
which is also accompanied by emission of radiation \cite{similar_calculations}.

We now turn to consider soliton propagation in a nonlinear optical waveguide 
with a periodic sequence of localized variations in the linear gain-loss with 
period $T$. In the reference frame, which is moving with velocity $2\beta_{0}$, 
the propagation is described by the following simple generalization of Eq. (\ref{cre7}): 
\begin{eqnarray}&&
i\partial_z\psi+\partial_t^2\psi+2|\psi|^2\psi=
i\epsilon_{1}\bar{c} \sum_{m=0}^{M} h\left(\bar{a}(t-(y_{b}+mT)+2\beta_{0}z)\right)\psi/2. 
\label{cre16}
\end{eqnarray}         
The distances at which the soliton encounters the gain-loss variations are 
$z_{cm}=z_{c1} + m\Delta z_{c}^{(1)}$, where $m=1, \dots, M$,  
$z_{c1}=(y_{b}-y_{0})/(2\beta_{0})$, and $\Delta z_{c}^{(1)}=T/(2|\beta_{0}|)$ 
is the inter-encounter distance. The rate of change of $\psi$ as a result of the 
order $\epsilon_{1}/|\beta_{0}|$ collisional effects, which are associated with 
the collisional frequency shift (\ref{cre15}), is given by 
$\exp\left[i\chi(t,z)\right]\Delta\Phi_{2}^{(1)}/\Delta z_{c}^{(1)}$. 
Therefore, the corresponding perturbation term in the NLS equation, which 
describes soliton propagation in the presence of a periodic sequence of localized 
variations in the linear gain-loss, is:   
\begin{eqnarray}&&
S=i\exp\left[i\chi(t,z)\right]
\frac{\Delta\Phi_{2}^{(1)}}{\Delta z_{c}^{(1)}}=
-\frac{\epsilon_{1}\bar{c} \tilde{h}}{2\bar{a}\beta_{0}T}
\exp\left[i\chi(t,z)\right]\partial_{t} |\psi_{s}(x)|.
\label{cre17}
\end{eqnarray}    
It follows that the perturbation term that is associated with the collision-induced 
frequency shift (\ref{cre15}) is indeed of the form $\epsilon_{s}\exp\left[i\chi(t,z)\right]
\partial_{t}|\psi|$, where $\epsilon_{s}=-(\epsilon_{1}\bar{c}\tilde{h})/[2\bar{a}\beta_{0}T]$.

\subsection{Derivation of a related perturbation term  
in waveguides with saturable cubic nonlinearity}
\label{model4}  

The derivation of the $\epsilon_{s}\exp\left[i\chi(t,z)\right]\partial_{t}|\psi|$ 
perturbation term in Subsection \ref{model3} is very useful from the physical point 
of view, since it shows the relevance of the term in systems other than optical 
fibers. However, from the mathematical point of view, the two derivations in 
Subsections \ref{model2} and \ref{model3} are closely related, 
as they both rely on the assumption of fast soliton interactions 
with weak perturbations. For this reason, it is useful to present another derivation 
of a related perturbation term, which is independent of the derivations in Subsections 
\ref{model2} and \ref{model3}. As we will see, this derivation is very helpful from 
the physical point of view as well, as it reveals a relation between the collisional 
Raman frequency shift term $\epsilon_{s}\exp\left[i\chi(t,z)\right]\partial_{t}|\psi|$ 
and the basic Raman self-frequency shift term $\epsilon_{R}\psi\partial_{t}|\psi|^2$, 
which describes the effects of delayed Raman response on single-pulse propagation.

We consider propagation of an optical pulse in a waveguide with second-order dispersion, 
cubic nonlinearity with quadratic saturation, and weak delayed Raman response. Due to 
the presence of quadratic saturation, the cubic nonlinearity term changes from 
$2|\psi|^2\psi$ to $2|\psi|^2\psi/(c_{1}+c_{2}|\psi|^2)$, where $c_{1}$ and $c_{2}$ 
are positive constants \cite{Pushkarov96,Herrmann92}. Additionally, since the basic 
delayed Raman response term $\epsilon_{R}\psi\partial_{t}|\psi|^2$ is a small 
correction to the cubic nonlinearity term \cite{Agrawal2019,Hasegawa87,Gordon86}, 
a similar change in the form should appear in the Raman term. 
That is, $\epsilon_{R}\psi\partial_{t}|\psi|^2$ should change 
to $\epsilon_{R}\psi\partial_{t}|\psi|^2/(c_{1}+c_{2}|\psi|^2)$. 
Therefore, the perturbed NLS equation that describes pulse propagation in the 
waveguide takes the form 
\begin{eqnarray}&&
i\partial_z\psi+\partial_t^2\psi+\frac{2|\psi|^2\psi}{c_{1}+c_{2}|\psi|^2}=
\frac{\epsilon_{R}\psi\partial_{t}|\psi|^2}{c_{1}+c_{2}|\psi|^2}. 
\label{cre21}
\end{eqnarray}   
We focus attention on the modified delayed Raman response term and notice that it can 
be written as $2\epsilon_{R}|\psi|^2\exp\left[i\chi(t,z)\right]\partial_{t}|\psi|
/(c_{1}+c_{2}|\psi|^2)$. In the limit of strong saturation, $c_{1}$ is a small 
parameter, which satisfies $0 < c_{1} \ll c_{2}|\psi|^2$ in some $t$ interval that 
includes the main body of the pulse. Therefore, in the limit of strong saturation, 
the modified delayed Raman response term can be approximated by 
\begin{eqnarray}&&
\frac{\epsilon_{R}\psi\partial_{t}|\psi|^2}{c_{1}+c_{2}|\psi|^2} \simeq 
\frac{2\epsilon_{R}}{c_{2}}\exp\left[i\chi(t,z)\right]\partial_{t}|\psi| 
\label{cre22}
\end{eqnarray}          
for $t$ and $z$, for which $0 < c_{1} \ll c_{2}|\psi|^2$. The right hand side of 
Eq. (\ref{cre22}) is of the form $\epsilon_{s}\exp\left[i\chi(t,z)\right]\partial_{t}|\psi|$  
with $\epsilon_{s}=2\epsilon_{R}/c_{2}$. Thus, we find that the collisional Raman frequency 
shift term has a form similar to the form attained by the modified delayed Raman response 
term near the main body of the pulse in waveguides with strong quadratic saturation of the 
cubic nonlinearity. We emphasize that away from the main body of the pulse, the inequality 
$c_{1} \ll c_{2}|\psi|^2$ breaks down even in the strong saturation limit. Therefore, a 
complete description of the effects of delayed Raman response on pulse propagation in 
a waveguide with quadratic saturation of the cubic nonlinearity is provided by the full 
modified Raman term $\epsilon_{R}\psi\partial_{t}|\psi|^2/(c_{1}+c_{2}|\psi|^2)$.

\section{Soliton dynamics in the presence of linear gain-loss, cubic loss, 
and the collisional Raman frequency shift}
\label{simu}

\subsection{Introduction}
\label{simu1} 

In two previous studies \cite{PC2018,PC2023}, we proposed and demonstrated a general mechanism 
for stabilizing the long-distance propagation of an optical soliton against radiation 
emission effects. The proposed stabilization mechanism was based on the interplay between 
frequency-dependent linear gain-loss and perturbation-induced shifting of the soliton's 
frequency. We demonstrated the stabilization for the following major nonlinear optical 
waveguide systems. (1) A waveguide system, where the frequency shift is caused by the 
effects of delayed Raman response on single-soliton propagation. (2) A waveguide system, 
where both the frequency shift and the frequency dependence of the gain-loss are due to 
guiding filters. The latter waveguide system was also studied in numerous earlier 
works \cite{Iannone98,Mollenauer2006,Mollenauer97,MM98,Mecozzi91,Mollenauer92,Kodama92,MMN96}. 
We showed that transmission stabilization in the two systems occurs via 
the same physical process. More specifically, the perturbation-induced 
frequency shift experienced by the soliton leads to the separation of the soliton 
and the radiation Fourier spectra, while the frequency-dependent linear gain-loss 
leads to efficient suppression of the emitted radiation.

In the current section, we investigate whether similar stabilization of soliton 
propagation occurs in the case where the dominant mechanism for the soliton's 
frequency shifting is due to the collisional Raman effect. More specifically, 
in Section \ref{simu2}, we demonstrate the strong radiative instability 
experienced by the soliton in waveguides with frequency-independent 
linear gain, cubic loss, and the collisional Raman frequency shift. 
Moreover, in Section \ref{simu3}, we investigate whether that strong radiative 
instability can be suppressed by the introduction of frequency-dependent linear 
gain-loss.

\subsection{Soliton dynamics in the presence of frequency-independent linear gain, 
cubic loss, and the collisional Raman frequency shift}
\label{simu2} 

We consider the propagation of an optical soliton in a nonlinear waveguide in the presence 
of weak frequency-independent linear gain, cubic loss, and the effects of the collisional 
Raman frequency shift. The propagation is described by the perturbed NLS equation (\ref{cre1}). 
Using the adiabatic perturbation theory for the NLS soliton \cite{Hasegawa95,PC2018,PC2020,Kaup90,CCDG2003}, 
we find the following equations for the dynamics of the soliton's amplitude 
and frequency \cite{P2007,CP2008,PC2012,PC2018}: 
\begin{equation}
\frac{d\eta}{dz}=\frac{4}{3}\epsilon_{3}\left(\eta_{0}^{2} - \eta^{2}\right)\eta,  
\label{cre25}
\end{equation} 
and 
\begin{equation}
\frac{d\beta}{dz} = -\frac{2}{3}\epsilon_{s}\eta^{2}, 
\label{cre26}
\end{equation} 
where $\eta_{0}$ is the amplitude value at stable equilibrium, 
and $g_{0}=4\epsilon_{3}\eta_{0}^{2}/3 > 0$ is used. Therefore, 
the $z$ dependences of the soliton's amplitude and frequency are given by: 
\begin{equation}
\eta(z) = \eta_{0}\left[1 + \left(\frac{\eta_{0}^{2}}{\eta^{2}(0)}-1\right)
\exp\left(-8\eta_{0}^{2}\epsilon_{3}z/3\right)\right]^{-\frac{1}{2}} , 
\label{cre27}
\end{equation}
and 
\begin{eqnarray}&&
\!\!\!\!\!\!\!\!\!
\beta(z) = \beta(0) 
-\frac{\epsilon_{s}}{4\epsilon_{3}}
\ln\left[1 - \frac{\eta^{2}(0)}{\eta_{0}^{2}} 
+\frac{\eta^{2}(0)}{\eta_{0}^{2}} \exp\left(8\epsilon_{3}\eta_{0}^{2}z/3\right)
\right].
\label{cre28}
\end{eqnarray}

The presence of linear gain is expected to lead to growth of the radiation emitted 
by the soliton, and to soliton destabilization. Similar radiative instability was 
observed in numerical simulations in Refs. \cite{PC2018,PC2023} for soliton propagation 
in the presence of frequency-independent linear gain and the effects of the Raman 
self-frequency shift. However, as we will see, the strength of the radiative instability 
and its characteristics in the current study are quite different from the ones observed 
in Refs. \cite{PC2018,PC2023}.

{\it Numerical simulations}.      
We solve Eq. (\ref{cre1}) numerically on a domain $[t_{min},t_{max}]
=[-400,400]$ using the split-step method with periodic boundary conditions 
\cite{Agrawal2019,Yang2010}. The initial condition is in the form of the single NLS 
soliton (\ref{cre2}) with amplitude $\eta(0)$, frequency $\beta(0)=0$, position $y(0)=0$, 
and phase $\alpha(0)=0$. We present here the simulations results for a typical case, 
where $\epsilon_{3}=0.01$, $\epsilon_{s}=0.032$, $\eta_{0}=1.0$, and $\eta(0)=0.8$, 
and emphasize that similar results are obtained for other values of the physical parameters. 
We note that the simulations setup used here is similar to the setup used in the simulations 
in Ref. \cite{PC2018}. The usage of similar setups enables a careful comparison between 
the results of numerical simulations with different propagation models \cite{epsilon_s_val}. 
The application of periodic boundary conditions means that the simulations describe 
propagation in a closed optical waveguide loop. Since many long-distance optical waveguide 
transmission experiments are carried out in closed loops \cite{Mollenauer2006,
Mollenauer97,MM98,Nakazawa2000,Nakazawa91}, the simulation setup that we use is very 
relevant from both the physical and the experimental points of view.

Pulse-shape distortion, transmission quality, and soliton stability are evaluated 
from the numerical simulations results by the transmission quality integrals 
$I^{(t)}(z)$ and $I^{(\omega)}(z)$, defined in Eqs. (\ref{Iz4}) and (\ref{Iz5}) 
in Appendix \ref{appendA}. $I^{(t)}(z)$ and $I^{(\omega)}(z)$ measure the deviations 
of the numerically obtained pulse shape $|\psi^{(num)}(t,z)|$ and Fourier spectrum 
$|\hat\psi^{(num)}(\omega,z)|$ from the perturbation method's predictions $|\psi^{(th)}(t,z)|$ 
and $|\hat\psi^{(th)}(\omega,z)|$, which are given by Eqs. (\ref{Iz1}) and (\ref{Iz3}), 
respectively. Therefore, $I^{(t)}(z)$ and $I^{(\omega)}(z)$ measure both distortions in the 
pulse shape and in the Fourier spectrum due to emission of radiation, and deviations in the 
numerically obtained values of the soliton's parameters from the perturbation method's 
predictions. Further insight into pulse-shape distortion and transmission quality 
is gained by measuring the distance $z_{q}$ at which the value of $I^{(t)}(z)$ first exceeds 
0.075. We refer to $z_{q}$ as the transmission quality distance. Additionally, pulse-shape 
distortion and soliton stability at larger distances are characterized by running the simulations 
up to a distance $z_{f}$, at which the value of $I(z)$ first exceeds 0.655 \cite{thresholds}. 
The values of $z_{q}$ and $z_{f}$ obtained in the numerical simulations with the parameter values 
specified in the preceding paragraph are $z_{q}=385$ and $z_{f}=695$.

\begin{figure}[ptb]
\begin{tabular}{cc}
\epsfxsize=8.1cm  \epsffile{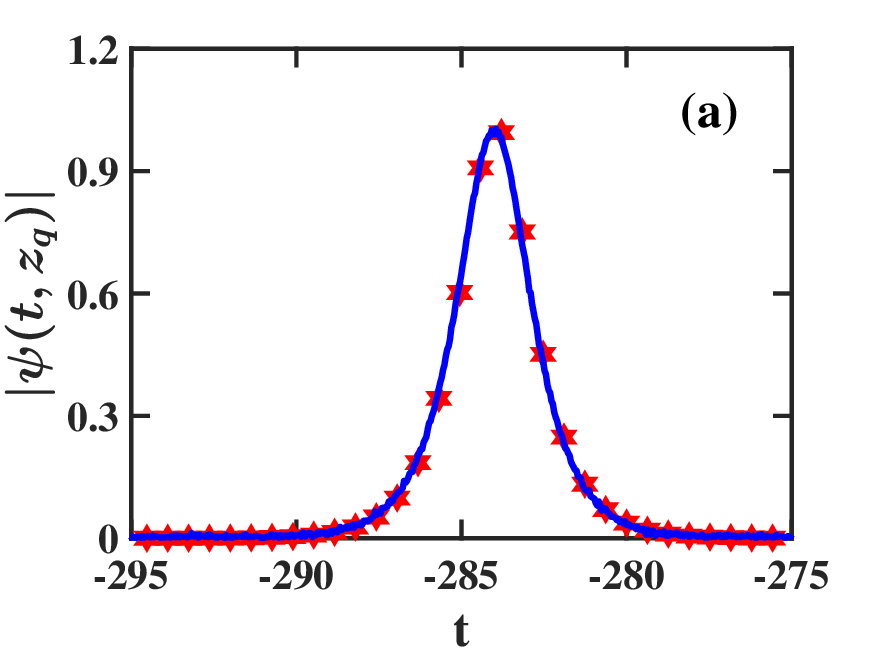} &
\epsfxsize=8.1cm  \epsffile{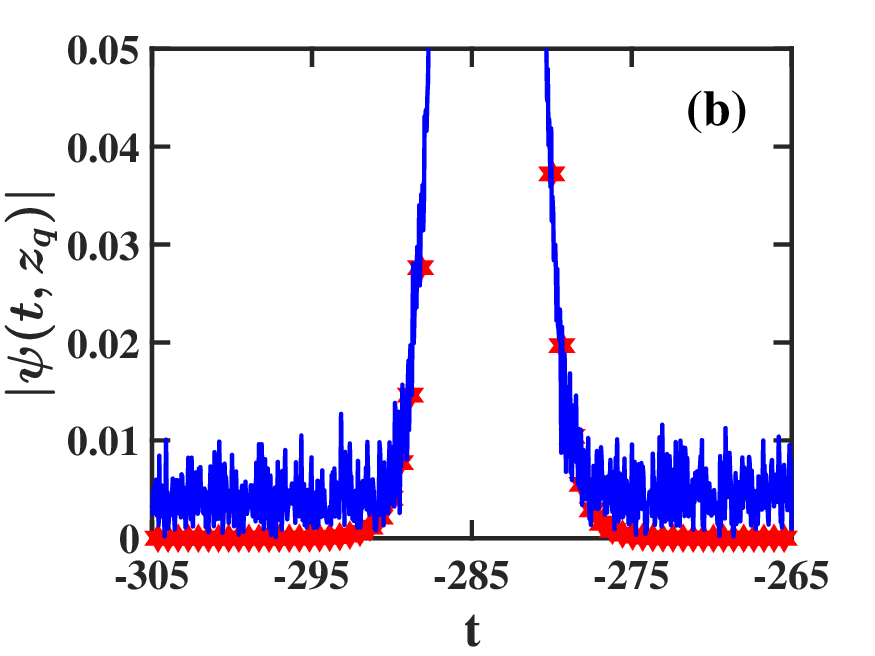} \\
\epsfxsize=8.1cm  \epsffile{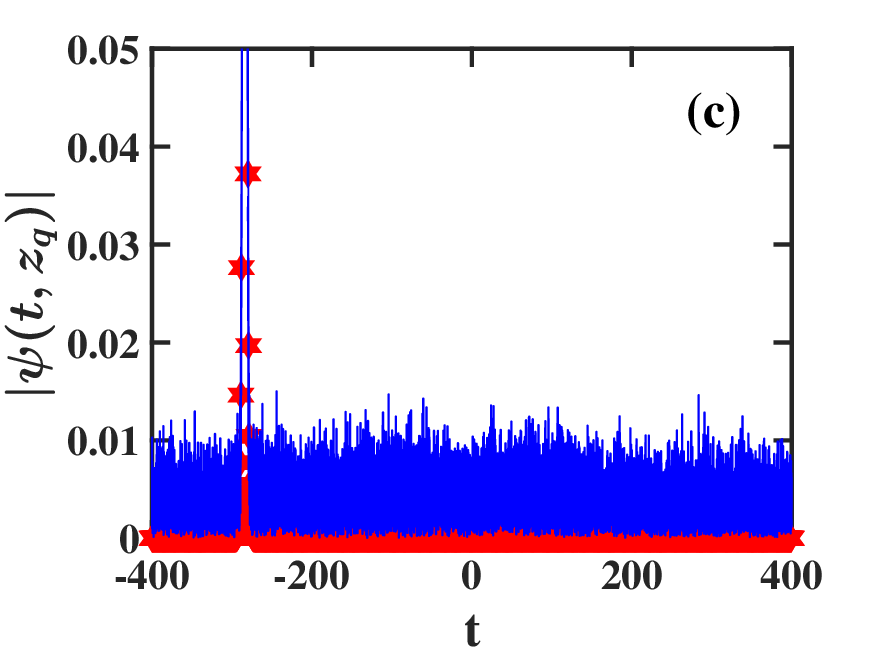} &
\epsfxsize=8.1cm  \epsffile{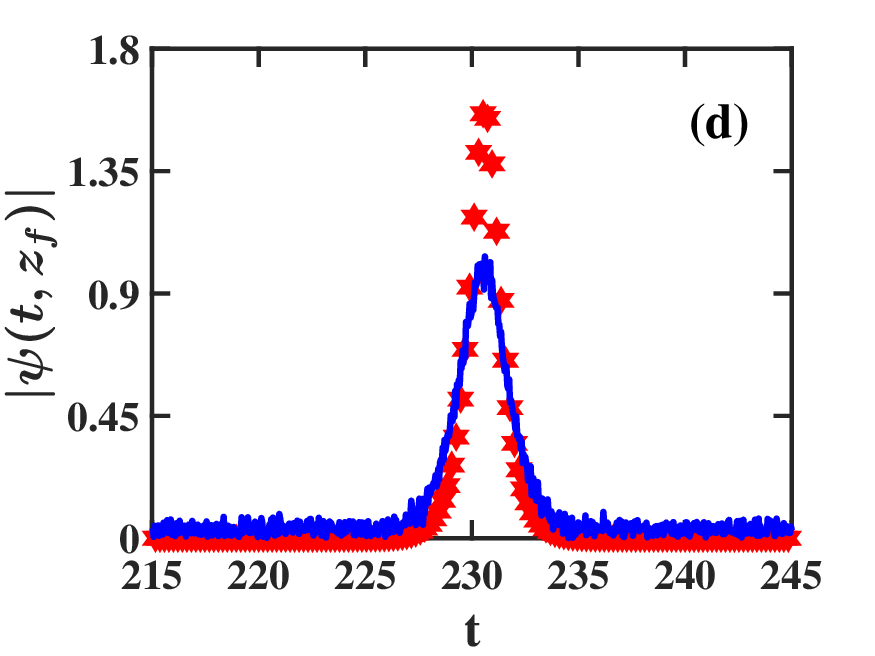} \\
\epsfxsize=8.1cm  \epsffile{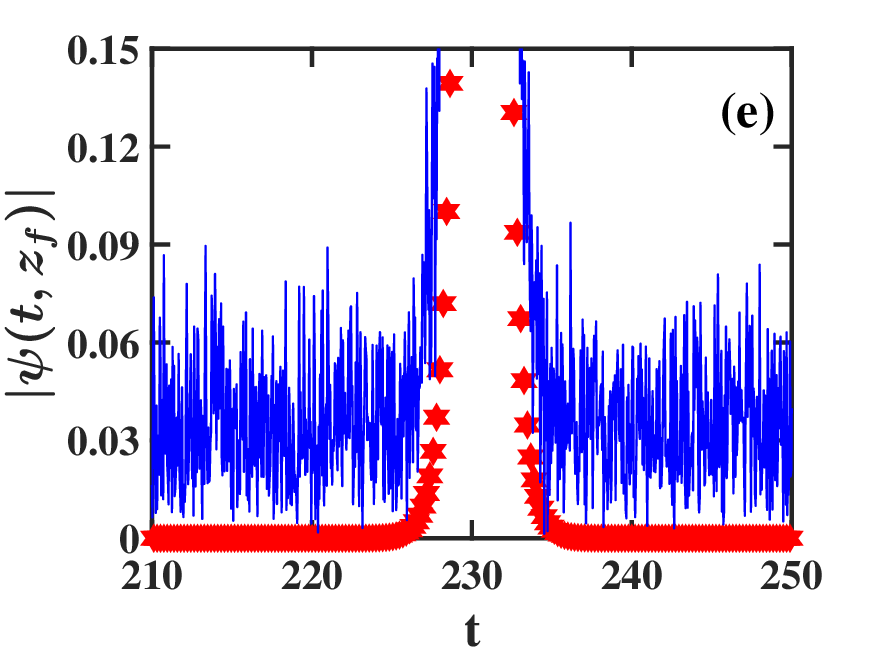} &
\epsfxsize=8.1cm  \epsffile{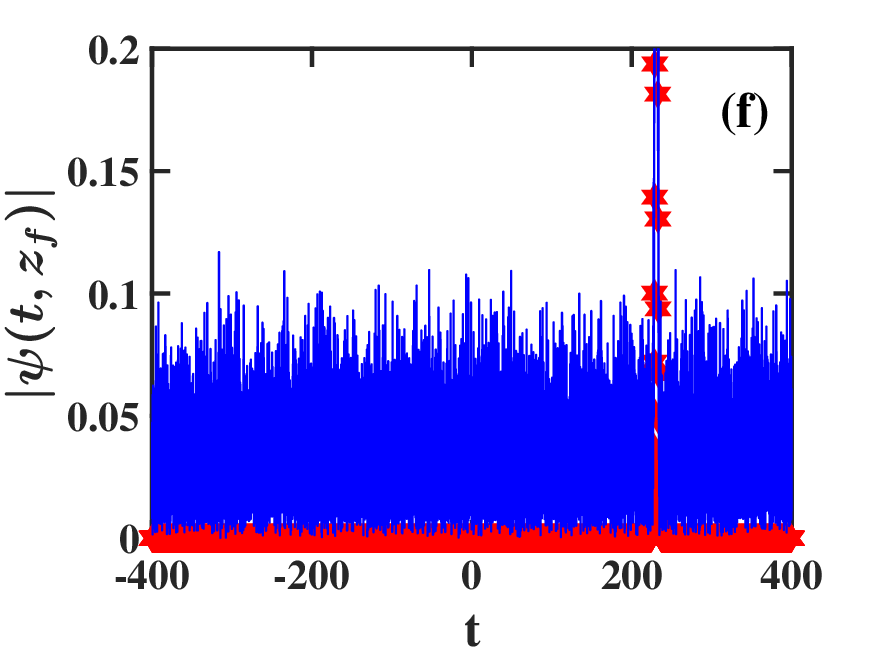}
\end{tabular}
\caption{The pulse shape $|\psi(t,z)|$ at $z_{q}=385$ [(a), (b), and (c)] and at 
$z_{f}=695$ [(d), (e), and (f)] for soliton propagation in a waveguide loop with weak 
frequency-independent linear gain, cubic loss, and the collisional Raman frequency shift. 
The physical parameter values are $\epsilon_{3}=0.01$, $\epsilon_{s}=0.032$, $\eta_{0}=1.0$, 
and $\eta(0)=0.8$. The solid blue curve represents the result obtained by numerical solution 
of Eq. (\ref{cre1}), and the red stars correspond to the perturbation method's prediction of 
Eqs. (\ref{Iz1}) and (\ref{cre27}).}
\label{fig1}
\end{figure}

\begin{figure}[ptb]
\begin{tabular}{cc}
\epsfxsize=8.1cm  \epsffile{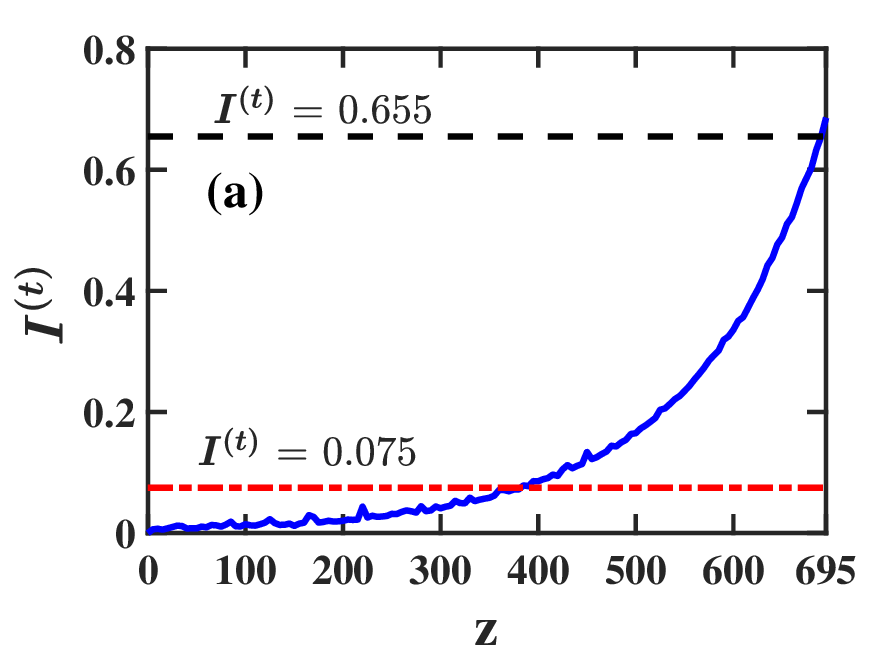} & 
\epsfxsize=8.1cm  \epsffile{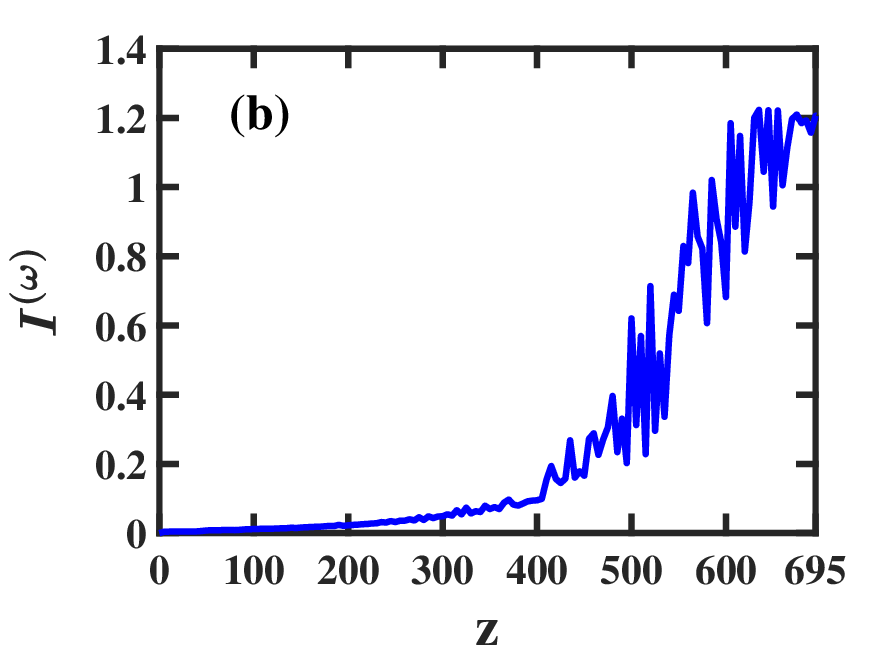} 
\end{tabular}
\caption{The $z$ dependences of the transmission quality integrals 
$I^{(t)}(z)$ [(a)] and $I^{(\omega)}(z)$ [(b)] obtained in the numerical simulation 
with Eq. (\ref{cre1}) for the same waveguide setup as in Fig. \ref{fig1}. The solid 
blue curves represent the simulation's results for $I^{(t)}(z)$ and $I^{(\omega)}(z)$. 
The dashed black and dashed-dotted red horizontal lines in (a) correspond to 
$I^{(t)}=0.655$ and $I^{(t)}=0.075$, respectively.}
\label{fig2}
\end{figure}

The pulse shape $|\psi(t,z)|$ obtained in the simulation at $z=z_{q}$ and at $z=z_{f}$ 
is shown in Fig. \ref{fig1} together with the perturbation method's prediction of 
Eqs. (\ref{Iz1}) and (\ref{cre27}). We observe that the numerically obtained pulse shape 
at $z=z_{q}$ is still close to the perturbation method's prediction, but that an appreciable 
radiative tail appears already at this distance [see Figs. \ref{fig1}(a)-\ref{fig1}(c)]. 
The radiative tail, which is generated by the three perturbation terms in Eq. (\ref{cre1}), 
consists of highly irregular and spiked oscillations, and is spread over the entire computational domain.  
As the soliton continues to propagate, the radiative tail keeps growing, and this leads to 
the significant pulse distortion that is observed in Figs. \ref{fig1}(d)-\ref{fig1}(f).  
Furthermore, the continued growth of the radiation-induced pulse distortion leads to the 
relatively rapid increase of the values of the transmission quality integrals with 
increasing $z$, which is seen in Fig. \ref{fig2}. It is important to note that the 
radiative tail observed in the current optical waveguide setup is significantly more 
irregular than the radiative tail that was observed in Ref. \cite{PC2018} for soliton 
propagation in the presence of the Raman self-frequency shift. The latter comparison 
indicates that the radiative instability induced by the collisional Raman frequency 
shift is stronger than the one induced by the Raman self-frequency shift. This 
conclusion is further supported by the smaller value of $z_{f}$ in the current 
simulation ($z_{f}=695$) compared with the $z_{f}$ value obtained in Ref. \cite{PC2018} 
with a similar simulation setup ($z_{f}=785$).

\begin{figure}[ptb]
\begin{center}
\begin{tabular}{cc}
\epsfxsize=8.2cm  \epsffile{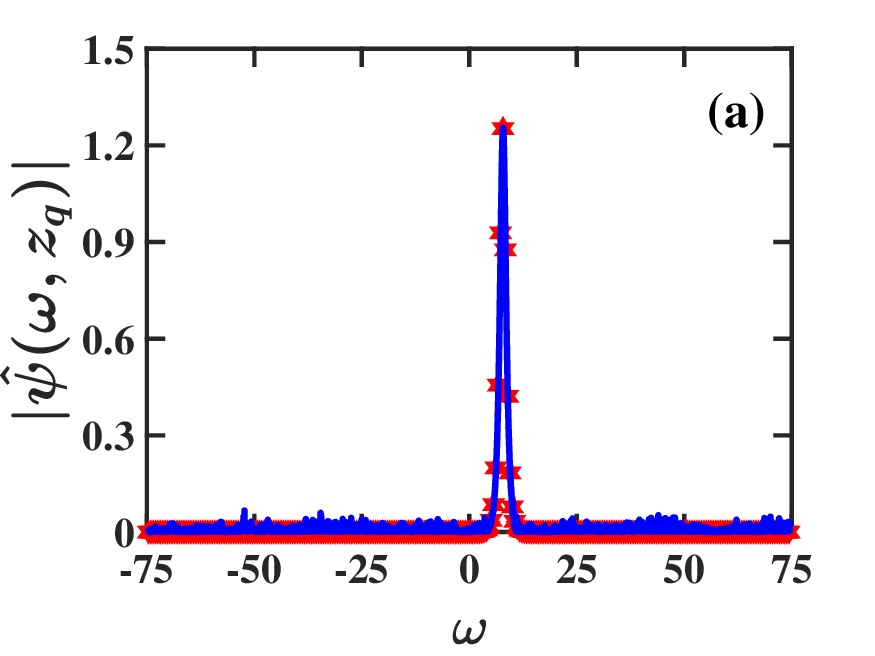} &
\epsfxsize=8.2cm  \epsffile{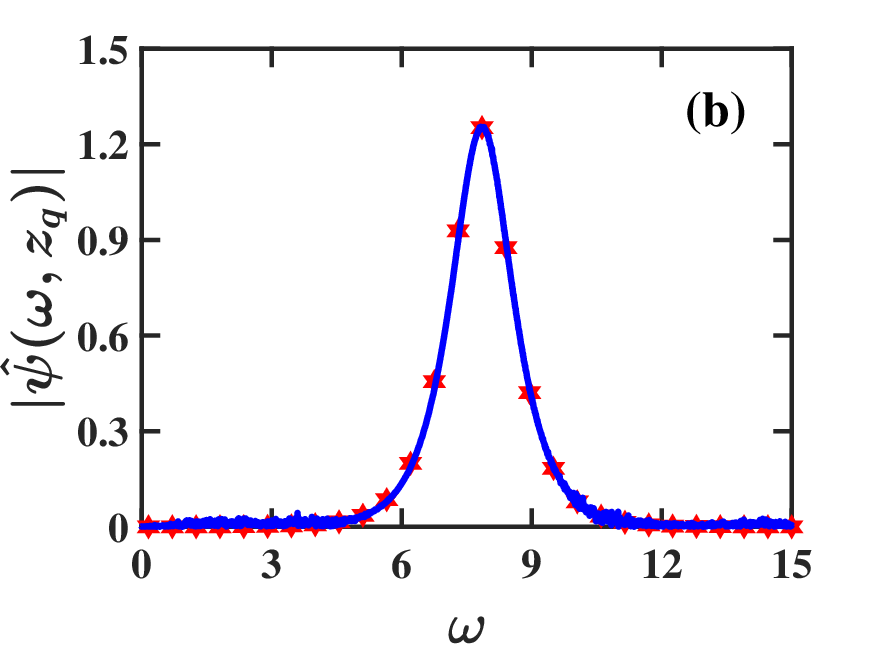} \\
\epsfxsize=8.2cm  \epsffile{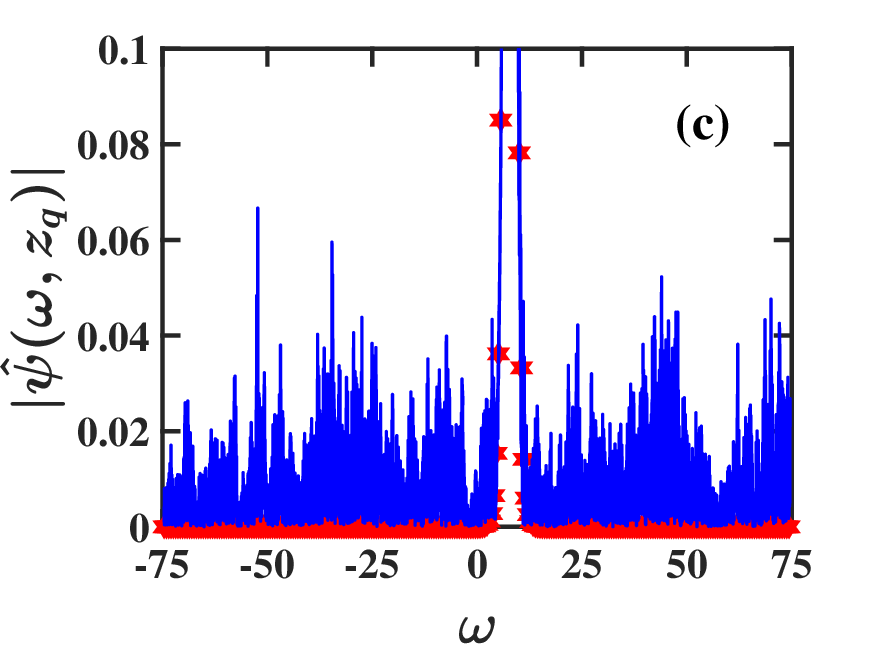} &
\epsfxsize=8.2cm  \epsffile{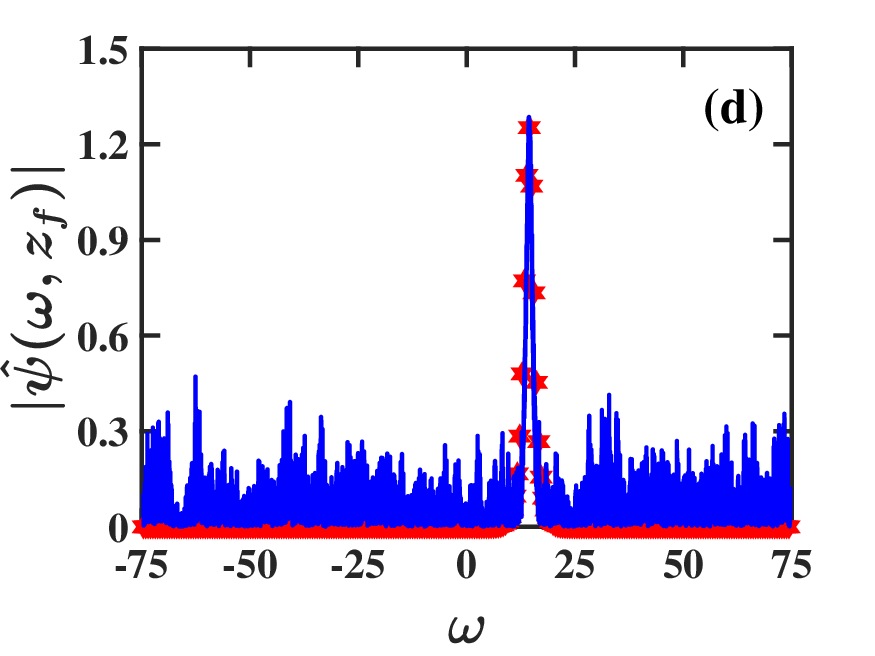} \\
\epsfxsize=8.2cm  \epsffile{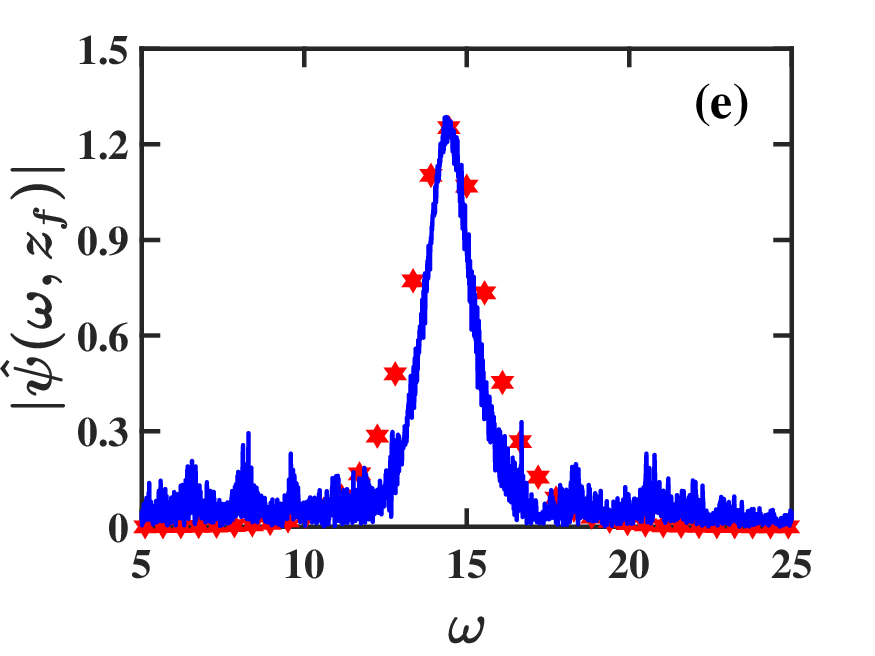}  
\end{tabular}
\end{center}
\caption{The Fourier spectrum of the optical pulse $|\hat\psi(\omega,z)|$ 
at $z_{q}=385$ [(a), (b), (c)] and at $z_{f}=695$ [(d), (e)] for soliton propagation 
in a waveguide loop with weak frequency-independent linear gain, cubic loss, and 
the collisional Raman frequency shift. The parameter values are the same as in 
Fig. \ref{fig1}. The solid blue curve represents the result obtained by numerical 
solution of Eq. (\ref{cre1}), and the red stars represent the perturbation method's 
prediction of Eqs. (\ref{Iz3}), (\ref{cre27}), and (\ref{cre28}).}                         
 \label{fig3}
\end{figure}

Further insight into pulse dynamics and soliton destabilization is obtained by 
analyzing the Fourier spectrum of the pulse $|\hat\psi(\omega,z)|$. Figure \ref{fig3} 
shows the numerically obtained Fourier spectrum at $z=z_{q}$ and at $z=z_{f}$ along with 
the perturbation method's prediction, obtained with Eqs. (\ref{Iz3}), (\ref{cre27}), 
and (\ref{cre28}). We observe that the soliton's Fourier spectrum is centered around 
the $z$ dependent soliton's frequency $\beta(z)$, and is partially separated from  
the radiation's spectrum. The radiation's spectrum is spread over the entire frequency 
interval used in the simulation $[-75,75]$, and is irregular and spiky. The partial separation 
of the soliton and radiation spectra is a result of the collisional Raman frequency shift 
experienced by the soliton. As a result of the partial spectral separation, the soliton 
part of the graph of $|\hat\psi^{(num)}(\omega,z)|$ is still close to the perturbation 
method's prediction at $z=z_{q}$ [see Fig. \ref{fig3}(b)].     
However, as the soliton continues to propagate, the radiative spectrum continues to grow 
and remains irregular [see Figs. \ref{fig3}(d)-\ref{fig3}(e)]. The growth of the radiative spectrum and 
its irregular nature lead to the strong pulse distortion that is observed in 
Figs. \ref{fig1} and \ref{fig2}.

We point out that the radiation spectrum observed in Fig. \ref{fig3} is drastically different 
from the radiation spectrum that was observed in Ref. \cite{PC2018} for soliton propagation 
in the presence of the Raman self-frequency shift. Indeed, the radiative spectrum in 
Fig. \ref{fig3} is very spiky and is spread over the entire frequency interval used in the 
simulation. In contrast, the radiative spectrum in Ref. \cite{PC2018} was rather smooth, 
and consisted of a single main peak, which was located near $\omega=0$ (see Fig. 7 in 
Ref. \cite{PC2018}). This comparison strengthens the conclusion drawn from Figs. \ref{fig1} 
and \ref{fig2} that the radiative instability generated by the collisional Raman frequency 
shift is stronger than the one generated by the Raman self-frequency shift.

\begin{figure}[ptb]
\begin{tabular}{cc}
\epsfxsize=8.1cm  \epsffile{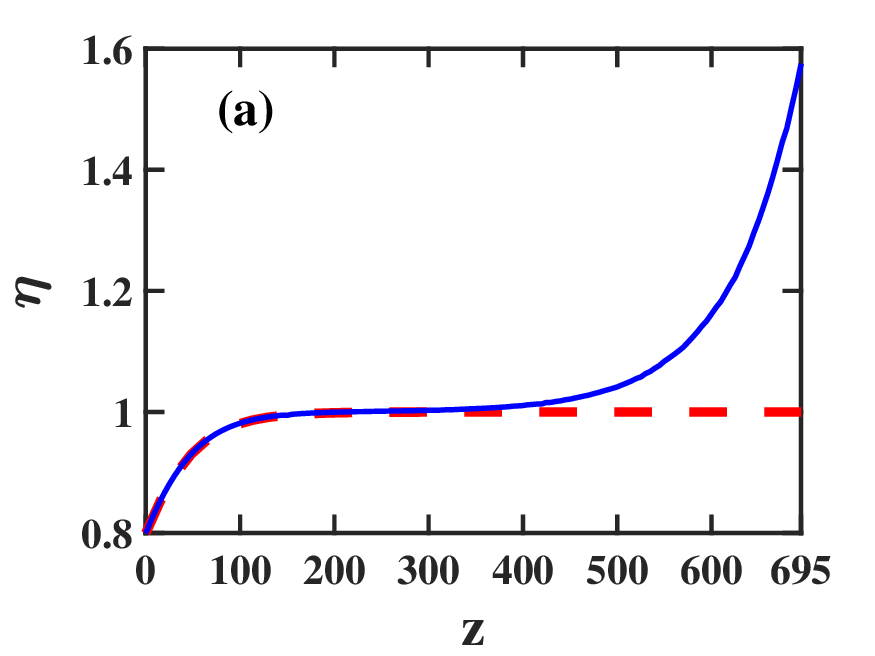} & 
\epsfxsize=8.1cm  \epsffile{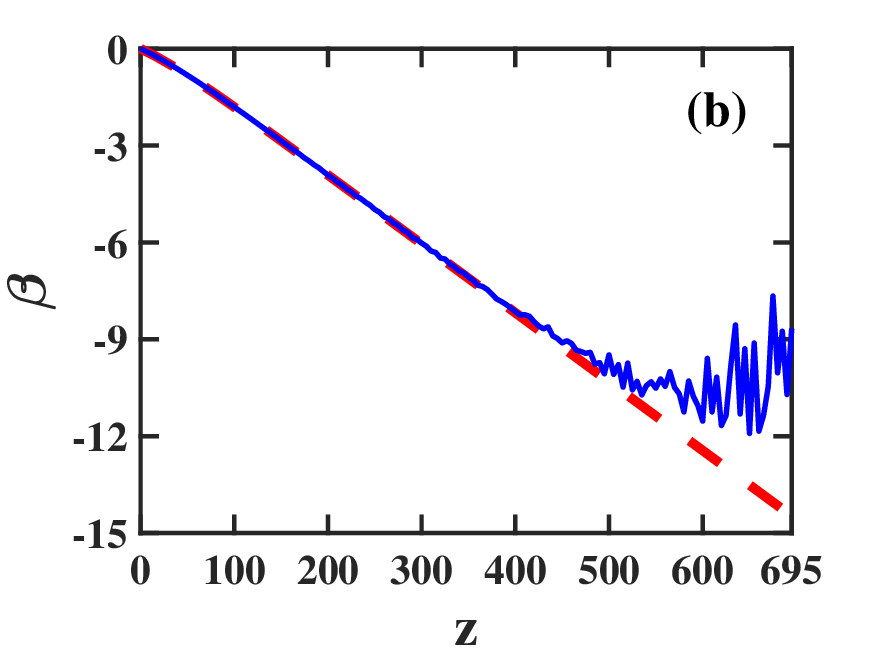}
\end{tabular}
\caption{The $z$ dependences of the soliton's amplitude $\eta(z)$ (a) 
and frequency $\beta(z)$ (b) for the same optical waveguide setup as in 
Figs. \ref{fig1}-\ref{fig3}. The solid blue curves represent the results 
obtained by numerical solution of Eq. (\ref{cre1}). The dashed red curves 
represent the perturbation theory predictions of Eq. (\ref{cre27}) in (a) 
and of Eq. (\ref{cre28}) in (b).}
\label{fig4}
\end{figure}

The destabilization of the NLS soliton in the presence of weak linear gain, cubic loss, 
and the collisional Raman frequency shift is also manifested in the dynamics of the 
soliton's amplitude and frequency. The $z$ dependences of the soliton's amplitude and 
frequency obtained in the simulation with Eq. (\ref{cre1}) are shown in Fig. \ref{fig4} 
together with the perturbation method's predictions of Eqs. (\ref{cre27}) 
and (\ref{cre28}). In both graphs we observe good agreement 
between the simulation's results and the predictions of Eqs. (\ref{cre27}) 
and (\ref{cre28}) for $0 \le z \le 400$. In contrast, for $400 < z \le 695$, 
the deviations of the numerical results from the predictions of Eqs. (\ref{cre27}) 
and (\ref{cre28}) become large. The increasing deviations coincide with the increase 
in pulse-shape and spectrum distortion that is observed in Figs. \ref{fig1} 
and \ref{fig3}. Based on the findings in Figs. \ref{fig1}-\ref{fig4} and on similar 
findings obtained with other physical parameter values, we conclude that soliton 
propagation in waveguide loops with weak frequency-independent linear gain, cubic loss, 
and the collisional Raman frequency shift is unstable. Similar radiative instabilities 
were observed in other waveguide setups with weak linear gain, cubic loss, and various 
perturbations that induce frequency shifts \cite{PC2018,PC2023}. However, as discussed 
in the preceding paragraphs, the radiative instability induced by the collisional Raman 
frequency shift appears to have some new characteristics and to be stronger than 
the instabilities in the waveguide setups considered in Refs. \cite{PC2018,PC2023}.

\subsection{Soliton dynamics in the presence of frequency-dependent linear gain-loss, 
cubic loss, and the collisional Raman frequency shift}
\label{simu3}

We now turn to address soliton propagation in a nonlinear optical waveguide in the 
presence of weak frequency-dependent linear gain-loss, cubic loss, and the collisional 
Raman frequency shift. Our main goal is to investigate whether the stronger 
radiative instability that was observed and characterized in Section \ref{simu2} 
can be suppressed by the interplay between frequency-dependent linear gain-loss 
and the collisional Raman frequency shift. Furthermore, we compare soliton 
stabilization in the current waveguide setup with soliton stabilization in 
the nonlinear waveguides with weak frequency-dependent linear gain-loss, cubic loss, 
and the Raman self-frequency shift, which were considered in Ref. \cite{PC2018}. 
Pulse propagation in the current nonlinear optical waveguide system is described 
by the following weakly perturbed NLS equation \cite{basic_model}:       
\begin{eqnarray} &&
i\partial_z\psi+\partial_t^2\psi+2|\psi|^2\psi=
i{\cal F}^{-1}(\hat g(\omega,z) \hat\psi)/2 - i\epsilon_{3}|\psi|^2\psi
+\epsilon_{s}\exp\left[i\chi(t,z)\right]\partial_{t}|\psi| , 
\label{cre31}
\end{eqnarray}    
where $\omega$ is angular frequency, $\hat\psi$ is the Fourier transform of $\psi$ 
with respect to time, $\hat g(\omega,z)$ is the frequency-dependent linear gain-loss,   
${\cal F}^{-1}$ is the inverse Fourier transform with respect to time, 
and the other notations are the same as in Eq. (\ref{cre1}).

The form of $\hat g(\omega,z)$ is chosen such that radiation emission effects are suppressed, 
soliton shape is preserved, and the value of the soliton's amplitude approaches the desired 
equilibrium value $\eta_{0}$. In particular, we choose the form 
\begin{eqnarray} &&
\!\!\!\!\!\!\!\!\!\!\!\!\!\!
\hat g(\omega,z) = -g_{L} + \frac{1}{2}\left(g_{0} + g_{L}\right)
\left[\tanh \left\lbrace \tilde\rho \left[\omega + \beta(z)+W/2\right] 
\right\rbrace 
- \tanh \left\lbrace \tilde\rho \left[\omega + \beta(z)- W/2\right] 
\right\rbrace\right],
\label{cre32}
\end{eqnarray} 
where $g_{0}$ is the linear gain coefficient, $\beta(z)$ is the soliton's frequency 
at distance $z$, and $g_{L}$ is an $O(1)$ positive coefficient. The constants $g_{0}$, 
$W$, and $\tilde\rho$ satisfy $0<g_{0}\ll 1$, $W \gg 1$, and $\tilde\rho\gg 1$. 
The form (\ref{cre32}) is similar to the one used in Refs. \cite{PC2018,PC2023} for 
soliton stabilization in the presence of the Raman self-frequency shift. The same 
form of $\hat g(\omega,z)$ was also used in studies of multisequence propagation of 
optical solitons in the presence of various weak dissipative perturbations 
\cite{PNT2016,PNH2017}. Since $\tilde\rho\gg 1$, $\hat g(\omega,z)$ can be 
approximated by the following step function: 
\begin{eqnarray} &&
\hat g(\omega,z) \simeq 
\left\{ \begin{array}{l l}
g_{0} &  \mbox{ if $-\beta(z)-W/2 < \omega < -\beta(z)+W/2$,}\\
(g_{0}-g_{L})/2 & \mbox{ if $\omega=-\beta(z) - W/2 \;\;$
or $\;\;\omega=-\beta(z) + W/2$,} \\ 
-g_{L} &  \mbox{elsewhere.}\\
\end{array} \right. 
\label{cre33}
\end{eqnarray}       
The possible advantages of the frequency-dependent linear gain-loss (\ref{cre32}) can 
be explained with the help of the approximation (\ref{cre33}). The weak linear gain 
$g_{0}$ in the frequency interval $(-\beta(z)-W/2, -\beta(z)+W/2)$ counteracts the effects 
of the weak cubic loss, such that the soliton's amplitude approaches the equilibrium value 
$\eta_{0}$ with increasing $z$. In addition, the relatively strong linear loss $-g_{L}$ diminishes 
emission of radiation with frequencies outside of the interval $(-\beta(z)-W/2, -\beta(z)+W/2)$. 
Moreover, the collisional Raman perturbation is expected to generate a frequency separation 
between the soliton's spectrum and the radiation's spectrum. Consequently, the introduction of the 
frequency-dependent linear gain-loss $\hat g(\omega,z)$ is expected to enable efficient 
suppression of pulse-shape distortion and significant enhancement of transmission stability.    
We note that the flat gain in the interval $(-\beta(z)-W/2, -\beta(z)+W/2)$ can be achieved   
by flat-gain amplifiers \cite{Becker99}, and the strong loss outside of this interval can be 
realized by filters \cite{Becker99} or by waveguide impurities \cite{Agrawal2019}.

We obtain the equations for the dynamics of the soliton's amplitude and frequency by employing  
the adiabatic perturbation method for the NLS soliton \cite{Hasegawa95,Kaup90,PC2018,PC2020,CCDG2003}. 
We find that amplitude dynamics is described by \cite{PC2018}: 
\begin{eqnarray} &&
\frac{d\eta}{dz} =
\left[-g_{L}  + \left( g_{0} + g_{L} \right) \tanh(V) - 4\epsilon_{3}\eta^{2}/3\right]\eta ,
\label{cre34}
\end{eqnarray}
where $V=\pi W/(4\eta)$. Additionally, the dynamics of the frequency is still described by 
Eq. (\ref{cre26}). Since we are interested in realizing stable soliton propagation with a 
constant positive amplitude $\eta_{0}$, we require that $\eta=\eta_{0}$ is a stable 
equilibrium point of Eq. (\ref{cre34}). This requirement yields the following equation 
for $g_{0}$: 
\begin{equation}
g_{0} = -g_{L} 
+\frac{\left(g_{L} + 4\epsilon_{3}\eta_{0}^{2}/3\right)}
{\tanh(V_{0})},   
\label{cre35}
\end{equation} 
where $V_{0}=\pi W/(4\eta_{0})$. Substituting  Eq. (\ref{cre35}) into  Eq. (\ref{cre34}), 
we arrive at \cite{PC2018}:    
\begin{eqnarray} &&
\frac{d\eta}{dz} = 
\left\lbrace g_{L}\left[\frac{\tanh(V)}{\tanh(V_{0})} - 1 \right]
+\frac{4}{3}\epsilon_{3}\left[\eta_{0}^{2}\frac{\tanh(V)}{\tanh(V_{0})} 
- \eta^{2} \right]\right\rbrace \eta.  
\label{cre36}
\end{eqnarray} 
Thus, the complete description of amplitude and frequency dynamics within the framework 
of the adiabatic perturbation method is given by Eqs. (\ref{cre36}) and (\ref{cre26}). 
One can show that the only equilibrium points of Eq. (\ref{cre36}) with a nonnegative 
amplitude value are $\eta=\eta_{0}$ and  $\eta=0$. Additionally,  $\eta=\eta_{0}$ is a 
stable equilibrium point, while $\eta=0$ is an unstable equilibrium point of the 
equation \cite{PC2018}.

{\it Numerical simulations}.      
The expectation that the interplay between the collisional Raman frequency shift and 
the weak frequency-dependent linear gain-loss would lead to stable long-distance 
soliton propagation is based on a heuristic argument. On one hand, a similar heuristic 
argument for stable soliton propagation in other waveguide setups was verified by 
numerical simulations in Refs. \cite{PC2018,PC2023}. On the other hand, as explained 
in Section \ref{simu2}, the radiative instability in waveguides with weak linear gain, 
cubic loss, and the collisional Raman frequency shift is stronger and more severe 
than the radiative instability in the waveguides with weak linear gain and cubic loss 
that were studied in Refs. \cite{PC2018,PC2023}. Therefore, it is important to check 
the validity of the heuristic argument for stable soliton propagation in the current 
waveguide setup by numerical simulations with the weakly perturbed NLS equation 
(\ref{cre31}) with the linear gain-loss (\ref{cre32}).

To enable comparison with the simulations results of Section \ref{simu2} and of Ref. 
\cite{PC2018}, we employ a simulation setup that is similar to the one used in Section \ref{simu2} 
and in Ref. \cite{PC2018}. More specifically, we solve Eqs. (\ref{cre31}) and (\ref{cre32}) 
numerically on a time domain $[t_{min},t_{max}]=[-400,400]$ with periodic boundary conditions 
by the split-step method \cite{Agrawal2019,Yang2010}. Since we use periodic boundary 
conditions, the simulation describes pulse propagation in a closed waveguide loop, 
which is the relevant propagation setup in many long-distance optical waveguide transmission 
experiments \cite{Mollenauer2006,Mollenauer97,MM98,Nakazawa2000,Nakazawa91}. The initial 
condition is in the form of the NLS soliton (\ref{cre2}) with parameter values $\eta(0)$, 
$\beta(0)=0$, $y(0)=0$, and $\alpha(0)=0$. Comparison with the results obtained in Section 
\ref{simu2} and in Ref. \cite{PC2018} is further enabled by choosing similar parameter 
values to the ones used in Section \ref{simu2} and in Ref. \cite{PC2018}. In particular, 
we use $\epsilon_{3}=0.01$, $\epsilon_{s}=0.032$, $\eta_{0}=1.0$, and $\eta(0)=0.8$, 
as well as $W=10$, $\tilde\rho=10$, and $g_{L}=0.5$ \cite{epsilon_s_val}. 
We stress that similar results to the ones presented in the current subsection are obtained with other 
physical parameters values. We run the numerical simulation up to a predetermined final distance 
$z_{f}=2000$, and monitor if the value of the transmission quality integral $I^{(t)}(z)$ 
remains smaller than 0.075 throughout the simulation.

\begin{figure}[ptb]
\begin{tabular}{cc}
\epsfxsize=8.1cm  \epsffile{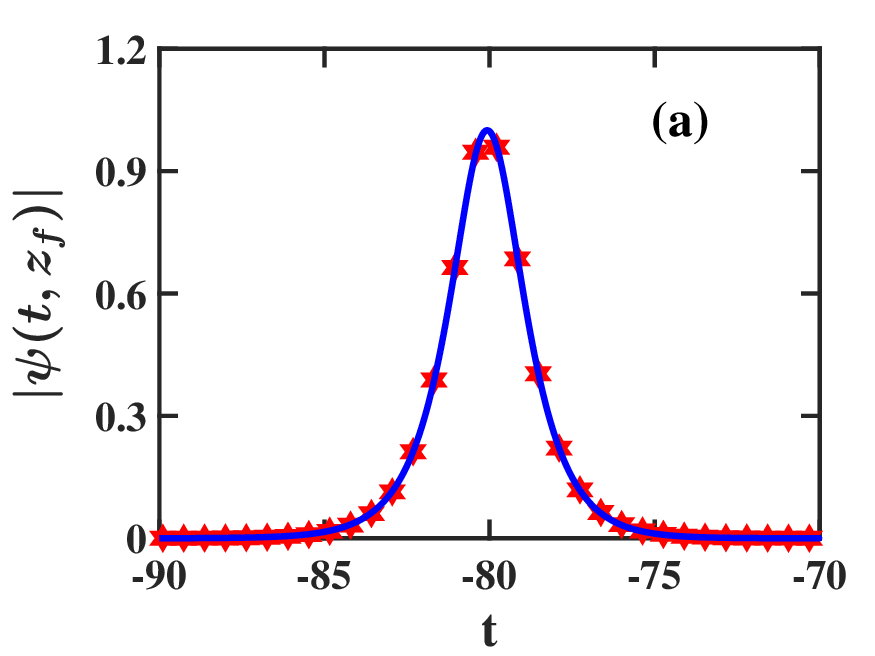} &
\epsfxsize=8.1cm  \epsffile{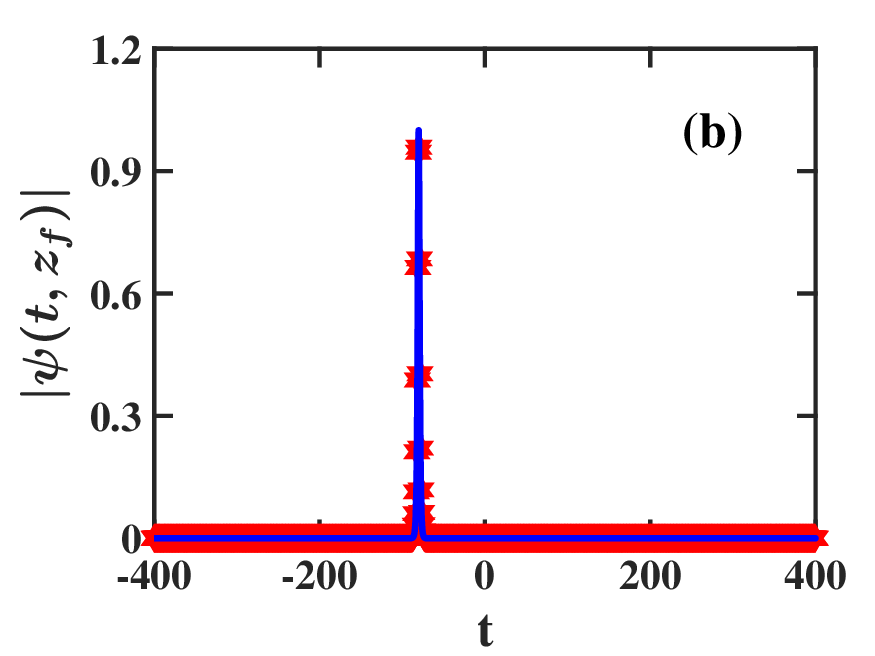} \\
\epsfxsize=8.1cm  \epsffile{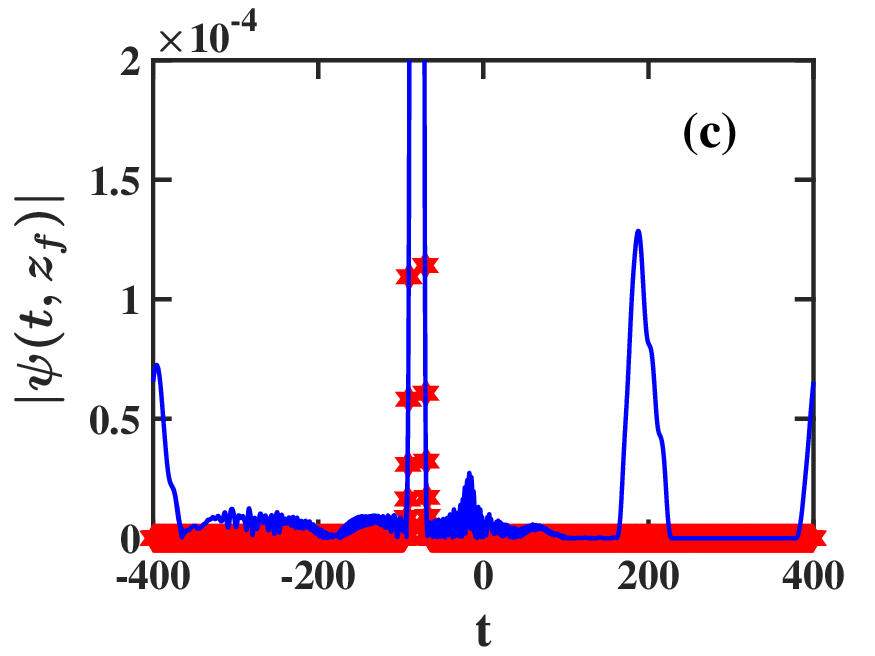} &
\epsfxsize=8.1cm  \epsffile{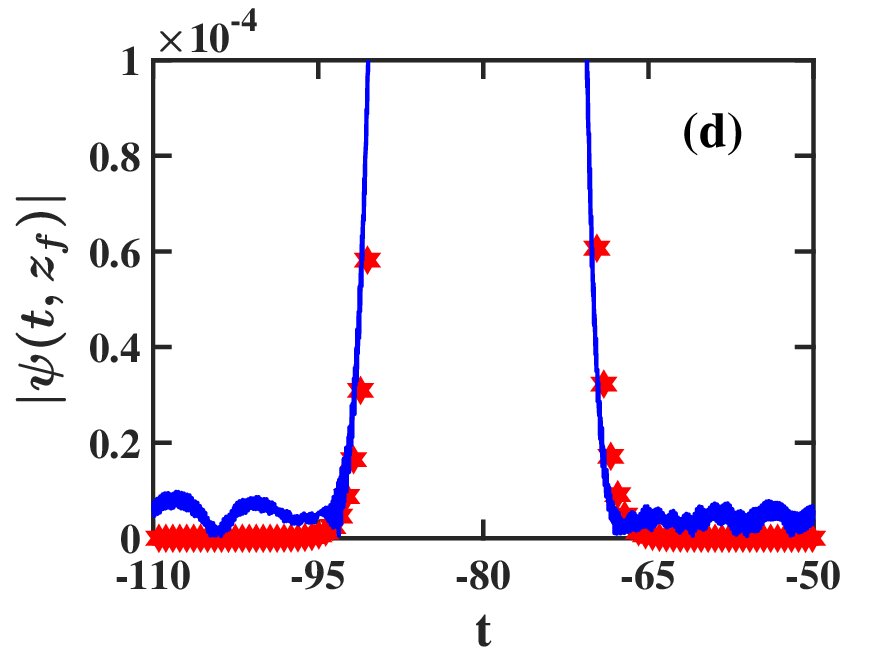} 
\end{tabular}
\caption{The pulse shape $|\psi(t,z_{f})|$,  where $z_{f}=2000$, for soliton 
propagation in a waveguide loop with weak frequency-dependent linear gain-loss, 
cubic loss, and the collisional Raman frequency shift. The physical parameter 
values are $\epsilon_{3}=0.01$, $\epsilon_{s}=0.032$, $\eta_{0}=1.0$, $\eta(0)=0.8$, 
$W=10$, $\tilde\rho=10$, and $g_{L}=0.5$. The solid blue curve corresponds to the 
result obtained by numerical solution of Eqs. (\ref{cre31}) and (\ref{cre32}).  
The red stars correspond to the perturbation method's prediction, obtained with 
Eqs. (\ref{Iz1}) and (\ref{cre36}).}
\label{fig5}
\end{figure}

\begin{figure}[ptb]
\begin{tabular}{cc}
\epsfxsize=8.1cm  \epsffile{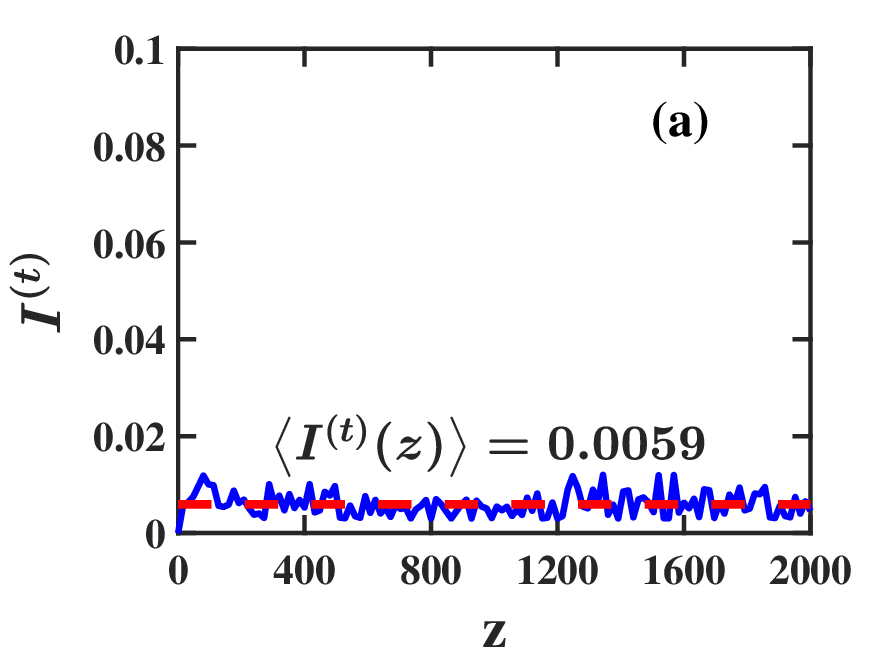} & 
\epsfxsize=8.1cm  \epsffile{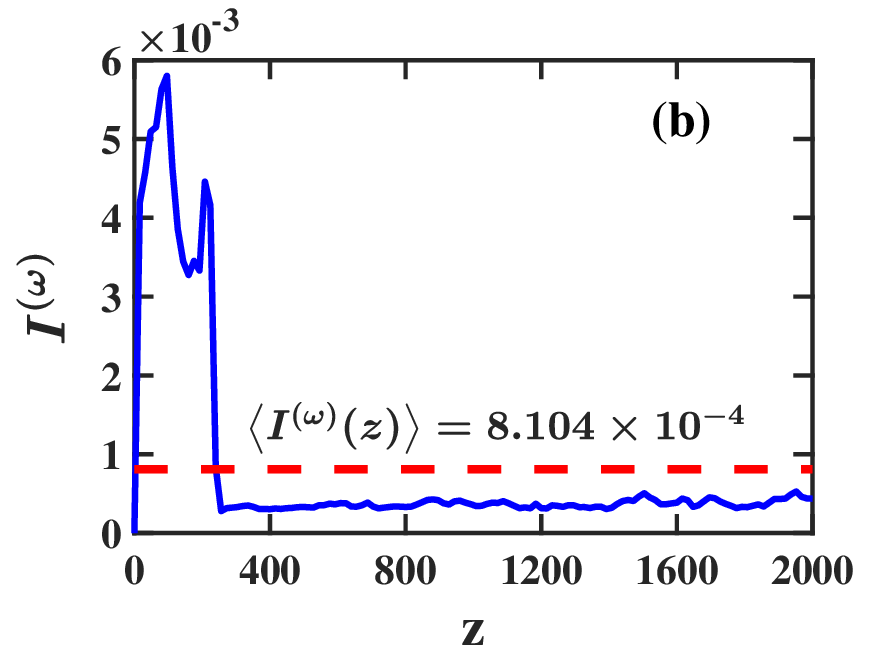} 
\end{tabular}
\caption{The $z$ dependences of the transmission quality integrals $I^{(t)}(z)$ [(a)] 
and $I^{(\omega)}(z)$ [(b)] obtained by the numerical simulation with Eqs. (\ref{cre31}) 
and (\ref{cre32}) for the same waveguide setup as in Fig. \ref{fig5}. The solid 
blue curves represent the simulation's results for $I^{(t)}(z)$ and $I^{(\omega)}(z)$. 
The dashed red horizontal lines correspond to $\langle I^{(t)}(z) \rangle$ and 
$\langle I^{(\omega)}(z) \rangle$.}
\label{fig6}
\end{figure}

The numerical simulation's result for the pulse shape $|\psi(t,z)|$ at $z=z_{f}$ 
is shown in Fig. \ref{fig5} along with the perturbation method's prediction of  
Eqs. (\ref{Iz1}) and (\ref{cre36}). As seen in Figs. \ref{fig5}(a) and  \ref{fig5}(b),   
the numerically obtained pulse shape is very close to the theoretical prediction. 
Additionally, the comparison of the theoretical and numerical results for small values 
of $|\psi(t,z_{f})|$ in Figs. \ref{fig5}(c) and \ref{fig5}(d) reveals that a rather 
weak radiative tail still exists at $z=z_{f}$. In particular, the values of 
$|\psi(t,z_{f})|$ in the radiative tail are smaller than $10^{-5}$ in the soliton's 
vicinity, and are smaller than $1.5 \times 10^{-4}$ in the entire temporal domain. 
Based on these findings and on similar results obtained with other physical parameter 
values we deduce that the interplay between weak frequency-dependent linear gain-loss 
and the collisional Raman frequency shift does lead to efficient mitigation of the 
strong pulse-shape distortion and the radiative instability, which were observed in 
Section \ref{simu2}. The enhancement of transmission stability is also evident in 
Fig. \ref{fig6}, which shows the numerically obtained curves of $I^{(t)}(z)$ and 
$I^{(\omega)}(z)$ together with the average values $\langle I^{(t)}(z) \rangle$ 
and $\langle I^{(\omega)}(z) \rangle$ \cite{averages}. As seen in the figure, the 
values of $I^{(t)}(z)$ and $I^{(\omega)}(z)$ remain smaller than 0.0120 and 0.0058 
throughout the propagation, in agreement with the small pulse distortion that is 
observed in Fig. \ref{fig5}. Accordingly, the average values are also small, 
$\langle I^{(t)}(z) \rangle = 0.0059$ and $\langle I^{(\omega)}(z) \rangle = 0.00081$. 
Furthermore, the value of $I^{(\omega)}(z)$ drops sharply inside a 
relatively small interval $(224.0,256.0)$ that contains the distance $z=251.1$ at 
which $|\beta(z)|$ first exceeds the value of $W/2$. Therefore, the sharp drop in 
the value of $I^{(\omega)}(z)$ occurs in the interval in which the soliton's and 
the radiation's Fourier spectra become separated.

\begin{figure}[ptb]
\begin{center}
\begin{tabular}{cc}
\epsfxsize=8.2cm  \epsffile{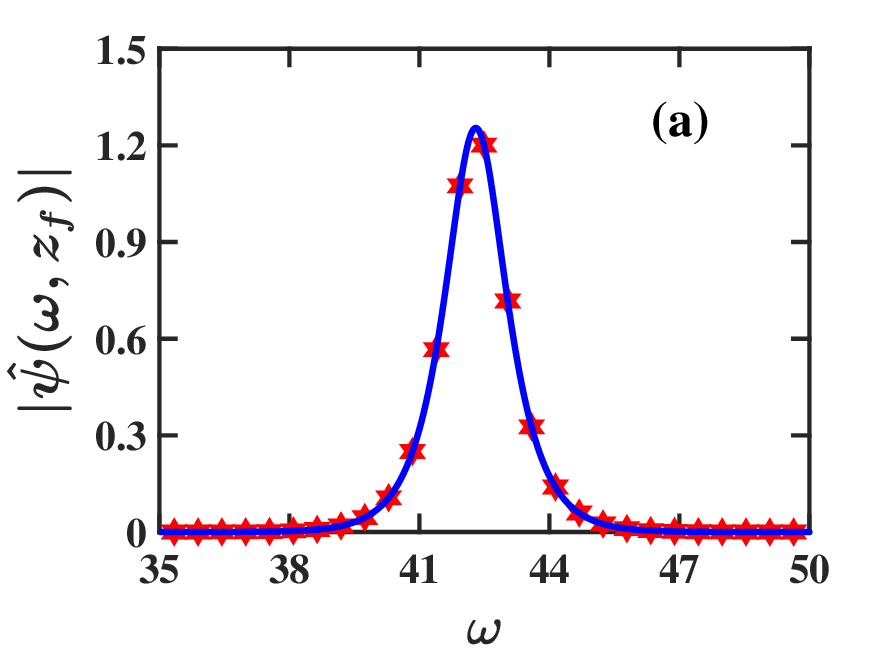} &
\epsfxsize=8.2cm  \epsffile{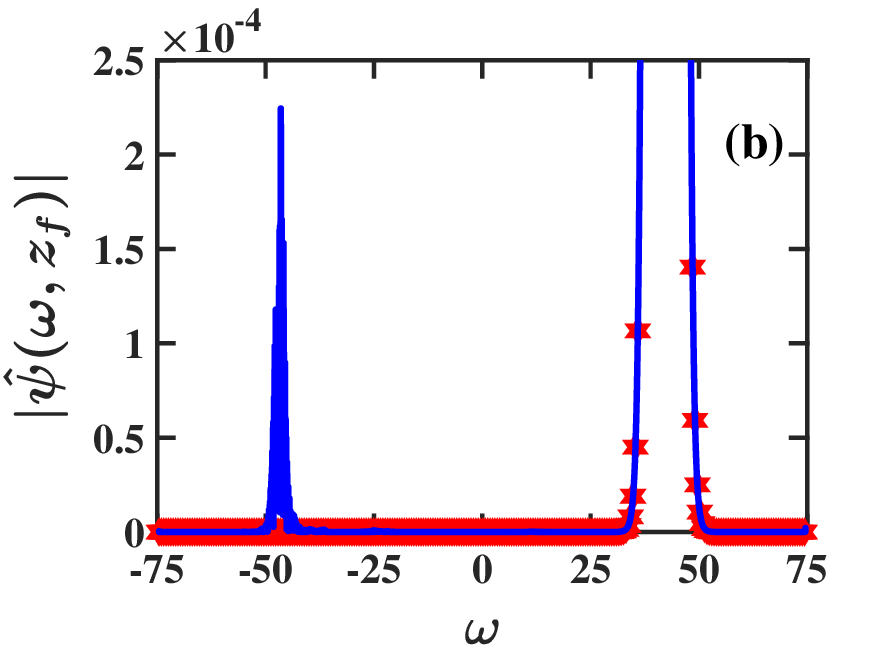} \\
\epsfxsize=8.2cm  \epsffile{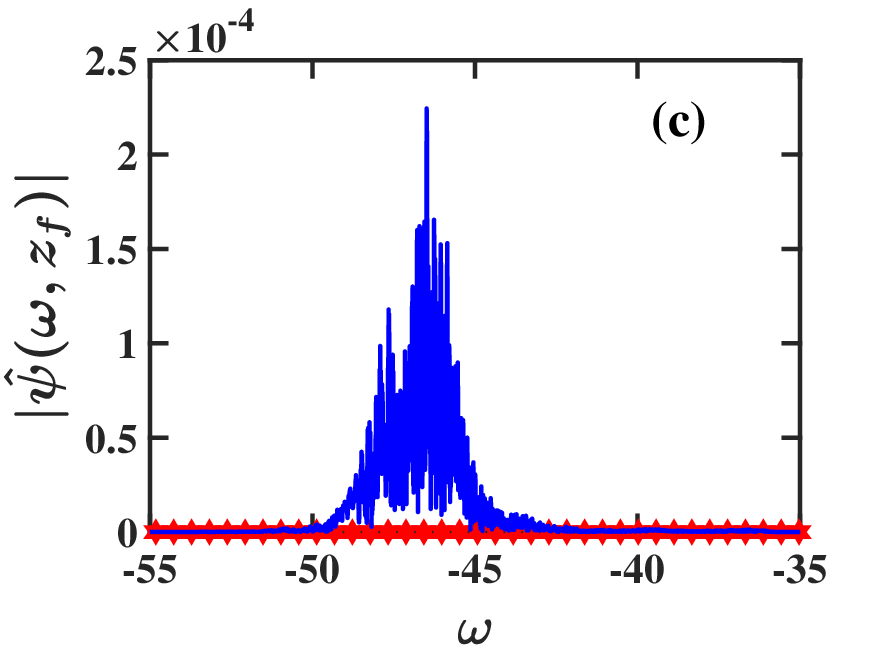} 
\end{tabular}
\end{center}
\caption{The Fourier spectrum of the optical pulse $|\hat\psi(\omega,z_{f})|$, where $z_{f}=2000$, 
for soliton propagation in a waveguide loop with weak frequency-dependent linear gain-loss, 
cubic loss, and the collisional Raman frequency shift. The parameter values are the same 
as in Fig. \ref{fig5}. The solid blue curve represents the result obtained by the numerical 
simulation with Eqs. (\ref{cre31}) and (\ref{cre32}). The red stars correspond to the 
perturbation method's prediction of Eqs. (\ref{Iz3}), (\ref{cre26}), and (\ref{cre36}).}                         
\label{fig7}
\end{figure}

Analysis of the Fourier spectrum of the pulse $|\hat\psi(\omega,z)|$ provides further 
insight into the enhancement of soliton stability. Figure \ref{fig7} shows a comparison 
between the numerically obtained Fourier spectrum at $z=z_{f}$ and the perturbation 
method's prediction of Eqs. (\ref{Iz3}), (\ref{cre26}), and (\ref{cre36}). The agreement 
between the theoretical and the numerical results is very good. More specifically, the 
theoretical and the numerical Fourier spectra of the soliton are both strongly downshifted,  
and are centered around the frequency $\omega_{m}=-\beta(z_{f})=42.30$. Furthermore, 
the graph of $|\hat\psi^{(num)}(\omega,z_{f})|$ vs $\omega$ does not contain significant 
spiky oscillations, such as the ones seen in Fig. \ref{fig3}, or significant radiation 
peaks such as the one observed in Ref. \cite{PC2018} for soliton propagation in waveguides 
with weak frequency-independent linear gain, cubic loss, and the Raman self-frequency shift.      
Based on these finding and on the results in Figs. \ref{fig1} and \ref{fig2}, we conclude 
that soliton transmission in the presence of weak frequency-dependent linear gain-loss, 
cubic loss, and the collisional Raman frequency shift is stable. Furthermore, 
despite the stronger radiative instability in the corresponding waveguide setup  
with weak linear gain, which was studied in Section \ref{simu2}, transmission 
stabilization occurs via the same stabilizing mechanisms that were observed in Refs. 
\cite{PC2018,PC2023}. That is, the effects of the collisional Raman frequency shift 
lead to separation of the soliton's Fourier spectrum from the radiation's Fourier spectrum, 
while the frequency-dependent linear gain-loss leads to effective mitigation of radiation emission.

We point out that the radiative effects in the current waveguide setup are stronger and have 
different characteristics compared with the ones observed in Ref. \cite{PC2018}, in stabilizing 
waveguide setups based on the Raman self-frequency shift. In particular, in the current waveguide 
setup, the strongest pulse-shape distortion features at $z_{f}=2000$ are of order $10^{-4}$. 
Additionally, these features appear as small shape-changing pulses that exist throughout the 
temporal domain. In contrast, in the waveguide setup considered in Section II B of Ref. \cite{PC2018}, 
the strongest pulse-shape distortion features at $z_{f}=2000$ were of order $10^{-7}$, appeared 
only in the soliton's vicinity, and attained the shapes of two distorted tails. Furthermore, 
the strongest spectrum distortion features in the current waveguide setup at $z_{f}=2000$ are 
of order $10^{-4}$, whereas the strongest spectral distortions in the waveguide setup 
considered in Section II B of Ref. \cite{PC2018} were of order $10^{-12}$. These findings 
strongly support our conclusion in Section \ref{simu2} of the current paper that radiation 
emission effects induced by the collisional Raman frequency shift are stronger that the ones 
generated by the Raman self-frequency shift. It is also interesting to compare the small 
radiation-induced pulses in the current waveguide setup with the pulse-shape distortions 
that were observed during transmission destabilization in the highly stable multisequence 
soliton-based transmission systems, which were studied in Section IV of Ref. \cite{PNT2016}.  
In the latter systems, stable multisequence transmission over distances exceeding $10^{4}$ 
dispersion lengths was achieved in nonlinear waveguides with weak frequency-dependent linear 
gain-loss and the collisional and noncollisional effects of delayed Raman response. 
Thus, we compare Fig. \ref{fig5}(c) in the current paper with Figs. 10(a) and 10(e) 
in Ref. \cite{PNT2016}. We observe that the small radiation-induced pulses in the current 
waveguide setup are similar in shape (but not in magnitude) 
to some of the most prominent radiation-induced pulse distortions, which appear 
during destabilization of multisequence soliton-based transmission at large distances. 
Therefore, this comparison strongly indicates that radiation emission due to the collisional 
Raman frequency shift plays an important role in transmission destabilization in the highly 
stable soliton-based multisequence transmission setups, which were considered in Section IV 
of Ref. \cite{PNT2016}.

\begin{figure}[ptb]
\begin{tabular}{cc}
\epsfxsize=8.1cm  \epsffile{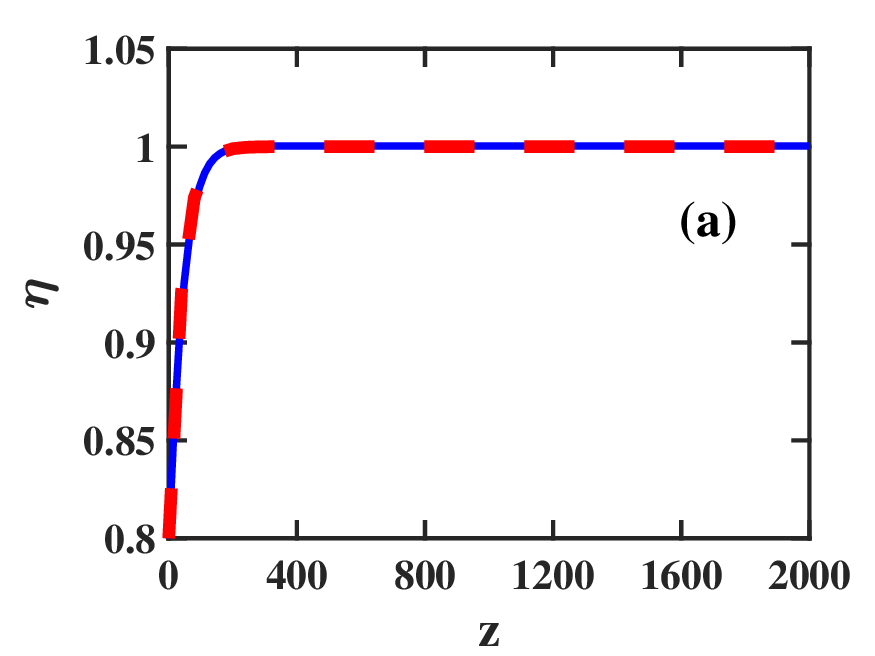} & 
\epsfxsize=8.1cm  \epsffile{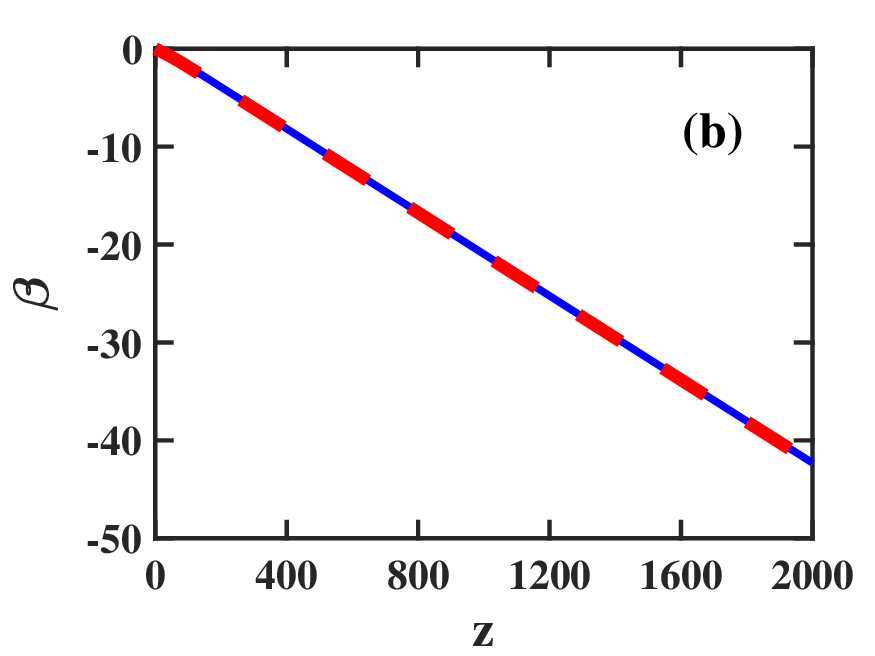}
\end{tabular}
\caption{The $z$ dependences of the soliton's amplitude $\eta(z)$ (a) 
and frequency $\beta(z)$ (b) for the same optical waveguide setup as in 
Figs. \ref{fig5}-\ref{fig7}. The solid blue curves represent the results 
obtained by the numerical simulation with Eqs. (\ref{cre31}) and (\ref{cre32}). 
The dashed red curves correspond to the perturbation method's predictions of 
Eq. (\ref{cre36}) in (a) and of Eqs. (\ref{cre26}) and (\ref{cre36}) in (b).}
\label{fig8}
\end{figure}

The enhancement of transmission stability in the presence of weak frequency-dependent 
linear gain-loss and the collisional Raman frequency shift has important implications 
for the dynamics of the soliton's amplitude and frequency. Figure \ref{fig8} shows the $z$ 
dependences of the soliton's amplitude and frequency obtained by the numerical simulation 
with Eqs. (\ref{cre31}) and (\ref{cre32}). Also shown are the perturbation method's 
predictions of Eqs. (\ref{cre26}) and  (\ref{cre36}). The agreement between the simulation's 
results and the perturbation method's predictions is excellent. More specifically, 
the numerically obtained value of $\eta(z)$ tends to the equilibrium value $\eta_{0}=1$ 
at short distances and remains close to this value throughout the propagation. Additionally, 
the numerically obtained value of $\beta(z)$ remains close to the $z$-dependent value 
predicted by the adiabatic perturbation method throughout the propagation. Therefore, 
the efficient suppression of radiation emission in the current waveguide setup due to 
the interplay between weak frequency-dependent linear gain-loss and the collisional 
Raman frequency shift enables realization of stable amplitude and frequency dynamics 
along significantly larger distances compared with the distances obtained with the 
waveguide setups of section \ref{simu2}, which were based on weak frequency-independent 
linear gain.

\section{Conclusions}
\label{conclusions}

We studied soliton stabilization against radiation emission in nonlinear optical 
waveguides in the presence of three perturbations due to weak linear gain-loss, cubic 
loss, and the collisional Raman frequency shift. We first explained how the collisional 
Raman frequency shift perturbation arises in the following three optical waveguide 
setups. (1) In multisequence soliton-based optical fiber transmission with a large 
number of soliton sequences. (2) In soliton propagation in nonlinear waveguides 
with weak localized variations in the linear gain or loss coefficient. 
(3) In soliton propagation in nonlinear waveguides with strong quadratic saturation 
of the cubic nonlinearity. We then investigated by numerical simulations with perturbed 
NLS models and by perturbative calculations the destabilization and stabilization of 
soliton transmission in the presence of the three aforementioned weak perturbations.

We began by considering soliton transmission in waveguides with weak frequency-independent 
linear gain, cubic loss, and the collisional Raman frequency shift. Our numerical simulations 
showed that in this case, the transmission is unstable due to emission of radiation.  
Furthermore, the radiative instability is stronger than the radiative instabilities that were 
observed in Ref. \cite{PC2018} for soliton propagation in the presence of weak linear gain, 
cubic loss, and various frequency-shifting physical processes. More specifically, the pulse 
distortions and the radiation's spectrum are significantly more spiky than the pulse distortions 
and the radiation's spectra that were found in Ref. \cite{PC2018}. Additionally, the radiation's 
spectrum has a substantially wider bandwidth compared with the radiation's spectra in Ref. \cite{PC2018}.  
As a result, in the current study, the separation between the soliton's and the radiation's spectra 
is partial, while in Ref. \cite{PC2018}, the spectral separation was full.

We then turned to investigate soliton transmission in waveguides with weak frequency-dependent 
linear gain-loss, cubic loss, and the collisional Raman frequency shift. In this case, our numerical 
simulations showed that the transmission is stable in spite of the stronger radiative instability 
in the corresponding waveguide setup with weak frequency-independent linear gain. Furthermore, 
we found that transmission stabilization is realized via the same generic mechanism that was 
proposed in Refs. \cite{PC2018,PC2023}. More precisely, the collisional Raman frequency shift 
experienced by the soliton leads to the separation of the soliton's and the radiation's Fourier 
spectra, and the frequency-dependent linear gain-loss leads to efficient suppression of 
radiation emission. Therefore, our study demonstrated the robustness and the general applicability 
of the proposed soliton stabilization method, which is based on the combined effects of 
perturbation-induced shifting of the soliton's frequency and frequency-dependent linear gain-loss. 
We also note that the weak pulse distortions observed in waveguides with weak frequency-dependent 
linear gain-loss, cubic loss, and the collisional Raman frequency shift are similar to 
the weak pulse-pattern distortions that were observed in Ref. \cite{PNT2016} 
at the onset of transmission instability of highly stable soliton-based multisequence transmission.     
This similarity indicates that radiation emission due to the collisional Raman frequency shift plays 
a key role in transmission destabilization in the highly stable soliton-based multisequence 
transmission setups, which were investigated in Ref. \cite{PNT2016}.

\section*{Acknowledgment}
D.C. is grateful to the Mathematics Department of NJCU 
for providing technological support for the numerical computations.

\appendix
\section{The transmission quality integrals $I^{(t)}(z)$ and $I^{(\omega)}(z)$} 
\label{appendA}

In this appendix, we present the perturbation method's predictions 
for the soliton's shape and Fourier spectrum, which are used in the 
analysis of transmission quality and transmission stability in 
Section \ref{simu}. Additionally, we present the method for calculating 
the transmission quality integrals $I^{(t)}(z)$ and $I^{(\omega)}(z)$ 
from the numerical simulations results.

The theoretical predictions for the soliton's shape and Fourier spectrum 
are based on the adiabatic perturbation method for the fundamental soliton 
of the NLS equation \cite{Hasegawa95,Kaup90,PC2018,PC2020,CCDG2003}. 
According to this method, one writes the solution $\psi(t,z)$ to the perturbed 
NLS equation as a sum $\psi(t,z)=\psi_{s}(t,z)+\psi_{rad}(t,z)$, 
where $\psi_{s}(t,z)$ is the soliton part, and $\psi_{rad}(t,z)$ is the 
radiation part \cite{Kaup90,PC2018,PC2020,CCDG2003}. The soliton part is given 
by the expression for the soliton solution to the unperturbed NLS equation with 
slowly varying parameters: $\psi_{s}(t,z)=\eta(z)\mbox{sech}(x)\exp(i\chi)$, 
where $x=\eta(z)\left[t-y(z)\right]$, $\chi(t,z)=\alpha(z)-\beta(z)\left[t-y(z)\right]$, 
$y(z)=y(0)-2\int_{0}^{z} dz' \beta(z')$, and $\alpha(z)=\alpha(0)+\int_{0}^{z} dz' \left[\eta^{2}(z')+\beta^{2}(z')\right]$ \cite{Kaup90,PC2018,PC2020,CCDG2003}. 
Similar to Refs. \cite{PC2018,PC2023}, we take $\psi_{s}(t,z)$ as the perturbation 
method's prediction for the soliton part: $\psi^{(th)}(t,z) \equiv \psi_{s}(t,z)$.  
Thus, the perturbation method's prediction for the soliton's shape is given by 
\begin{eqnarray} 
|\psi^{(th)}(t,z)|=\eta(z)\mbox{sech}\left[\eta(z)\left(t-y(z)\right)\right],
\label{Iz1}
\end{eqnarray}  
where $\eta(z)$ is calculated by solving the perturbation method's equation for 
$d\eta/dz$, and $y(z)$ is measured from the simulations \cite{PC2018,PC2023}.  
The Fourier transform of $\psi_{s}(t,z)$ with respect to time is 
\begin{eqnarray} 
\hat\psi_{s}(\omega,z)=
\left(\frac{\pi}{2}\right)^{1/2}
\frac{\exp[i\alpha(z)-i\omega y(z)]}
{\cosh\left[\pi\left(\omega+\beta(z)\right)/\left(2\eta(z)\right)\right]}.
\label{Iz2}
\end{eqnarray}               
Therefore, the theoretical prediction for the Fourier spectrum of the soliton is
\begin{eqnarray} 
\!\!\!\!\!\!\!
|\hat\psi^{(th)}(\omega,z)|=
\left(\frac{\pi}{2}\right)^{1/2}
\mbox{sech}\left[\pi\left(\omega+\beta(z)\right)/\left(2\eta(z)\right)\right], 
\label{Iz3}
\end{eqnarray}   
where $\eta(z)$ and $\beta(z)$ are calculated by solving the perturbation method's 
equations for $d\eta/dz$ and $d\beta/dz$.

The transmission quality integrals $I^{(t)}(z)$ and $I^{(\omega)}(z)$ measure the deviations 
of the numerically obtained pulse shape $|\psi^{(num)}(t,z)|$ and Fourier spectrum 
$|\hat\psi^{(num)}(\omega,z)|$ from the perturbation theory's predictions $|\psi^{(th)}(t,z)|$ 
and $|\hat\psi^{(th)}(\omega,z)|$. We define $I^{(t)}(z)$ in the same way as we did in 
previous studies of soliton propagation in nonlinear optical waveguide systems 
\cite{PC2018,PC2023,PNT2016}: 
\begin{eqnarray} &&
\!\!\!\!\!\!\!\!\!\!
I^{(t)}(z)=   
\left[ \int_{t_{min}}^{t_{max}} dt\,
\left| \psi^{(th)}(t,z) \right|^2 \right]^{-1/2}
\left\{\int_{t_{min}}^{t_{max}} \!\!\!\!\!\!\!\!\! dt\,
\left[\;\left|\psi^{(th)}(t,z) \right| - 
\left|\psi^{(num)}(t,z) \right| \; \right]^2 \right\}^{1/2}
\!\!\!\!,  
\label{Iz4}
\end{eqnarray}  
where the integration is performed over the entire time domain used in the simulation 
$[t_{min},t_{max}]$. $I^{(\omega)}(z)$ is defined in a similar manner by:      
\begin{eqnarray} &&
\!\!\!\!\!\!\!\!\!\!\!\!\!\!\!
I^{(\omega)}(z)=   
\left[ \int_{\omega_{min}}^{\omega_{max}} d\omega\,
\left| \hat{\psi}^{(th)}(\omega,z) \right|^2 \right]^{-1/2}
\left\{\int_{\omega_{min}}^{\omega_{max}} \!\!\!\!\!\!\!\!\! d\omega\,
\left[\;\left|\hat{\psi}^{(th)}(\omega,z) \right| - 
\left|\hat{\psi}^{(num)}(\omega,z) \right| \; \right]^2 \right\}^{1/2}
\!\!\!\!,  
\label{Iz5}
\end{eqnarray}  
where the integration interval $[\omega_{min},\omega_{max}]$ is the entire frequency 
interval used in the simulation. From these definitions it is clear that $I^{(t)}(z)$ 
and $I^{(\omega)}(z)$ measure both distortions in the pulse shape and in the Fourier spectrum 
due to radiation emission, and deviations of the numerically obtained values of the soliton's 
amplitude and frequency from the values predicted by the perturbation method.  
The transmission quality distance $z_{q}$ is defined as the distance at which the value of 
$I^{(t)}(z)$ first exceeds a constant value $K$. In the current paper we used $K=0.075$. 
We point out that the values of $z_{q}$ obtained by using this definition are not very 
sensitive to the value of $K$, in the sense that small changes in the value of $K$ lead to 
small changes in the measured $z_{q}$ values.


\begin{thebibliography}{}
\bibitem{epsilon_s_val} The $\epsilon_{s}$ value is taken as the value that yields the same 
frequency shift for a unit amplitude soliton [$\Delta\beta^{(c)}(z)=-2\epsilon_{s}z/3$] 
as the frequency shift experienced by a unit amplitude soliton due to the Raman self-frequency 
shift [$\Delta\beta^{(s)}(z)=-8\epsilon_{R}z/15$]. For the value of $\epsilon_{R}=0.04$ that 
was used in Ref. \cite{PC2018} we find $\epsilon_{s}=0.032$.
\bibitem{averages} The averages $\langle I^{(t)}(z) \rangle$ and 
$\langle I^{(\omega)}(z) \rangle$ are defined by: 
$\langle I^{(t)}(z) \rangle = \int_{0}^{z_{f}} dz' I^{(t)}(z') /z_{f}$ and 
$\langle I^{(\omega)}(z) \rangle = \int_{0}^{z_{f}} dz' I^{(\omega)}(z') /z_{f}$, 
respectively.
\end{thebibliography}
\end{document}